\documentclass[aps,
superscriptaddress,
showpacs,
letterpaper,
twocolumn,pra]{revtex4}
\newcommand{\expec}[1]{\langle #1\rangle}

\usepackage{linearb}
\usepackage{amsmath}
\usepackage{amsfonts}
\usepackage{enumitem}
\usepackage{grffile}
\usepackage{graphicx}
\usepackage{amsthm}
\usepackage{subfigure}
\usepackage{array}
\usepackage{hyperref}
\usepackage[usenames]{color}
\usepackage{slashbox}

\usepackage{bbold}

\usepackage{multirow}

\usepackage{braket}

\renewcommand{\v}[1]{\mathbf{#1}}

\newcommand{\lp}{\left ( }
\newcommand{\rp}{\right ) }
\newcommand{\lb}{\left [ }
\newcommand{\rb}{\right ] }

\newcommand{\beq}{\begin{eqnarray*}}
\newcommand{\eeq}{\end{eqnarray*}}

\newcommand{\be}{\begin{eqnarray}}
\newcommand{\ee}{\end{eqnarray}}

\newcommand{\mc}{\mathcal}

\newcolumntype{M}{>{\centering\arraybackslash} m }

\def\lsim{\mathrel{\rlap{\lower4pt\hbox{\hskip1pt$\sim$}}
    \raise1pt\hbox{$<$}}}                

\def\gsim{\mathrel{\rlap{\lower4pt\hbox{\hskip1pt$\sim$}}
    \raise1pt\hbox{$>$}}}                

\definecolor{salvatore}{rgb}{0,0,1}

\begin{document}

\title{Quantum correlations and entanglement in far-from-equilibrium spin systems}
\author{Kaden R.~A. Hazzard} \email{kaden.hazzard@gmail.com}
\affiliation{JILA and Department of Physics, University of Colorado-Boulder, NIST, Boulder, Colorado 80309-0440, USA}
\author{Mauritz van den Worm}
\affiliation{Institute of Theoretical Physics, University of Stellenbosch, Stellenbosch 7600, South Africa}
\author{Michael Foss-Feig}
\affiliation{JQI, NIST, Department of Physics, University of Maryland, College Park, MD 20742, USA}
\author{Salvatore R. Manmana}
\affiliation{Institute for Theoretical Physics, Georg-August-Universit\"at G{\"o}ttingen, D-37077 G{\"o}ttingen, Germany}
\author{Emanuele Dalla Torre}
\affiliation{Department of Physics,
Harvard University, Cambridge, MA 01238, USA}
\author{Tilman Pfau}
\affiliation{5. Physikalisches Institut and Center for Integrated Quantum Science and Technology, Universit{\"a}t Stuttgart, Pfaffenwaldring 57, 70569 Stuttgart, Germany
}
\author{Michael Kastner}
\affiliation{National Institute for Theoretical Physics (NITheP), Stellenbosch 7600, South Africa}
\affiliation{Institute of Theoretical Physics, University of Stellenbosch, Stellenbosch 7600, South Africa}
\author{Ana Maria Rey}
\affiliation{JILA and Department of Physics, University of Colorado-Boulder, NIST, Boulder, Colorado 80309-0440, USA}

\begin{abstract}

By applying complementary analytic and numerical methods, we investigate the dynamics of spin-$1/2$ XXZ models with variable-range interactions in arbitrary dimensions. The dynamics we consider is initiated from uncorrelated states that are easily prepared in experiments, and can be equivalently viewed as either Ramsey spectroscopy or a quantum quench.  Our primary focus is the dynamical emergence of correlations and entanglement in these far-from-equilibrium interacting quantum systems: we characterize these correlations by the entanglement entropy, concurrence, and squeezing, which are inequivalent measures of entanglement corresponding to different quantum resources.  In one spatial dimension, we show that the time evolution of correlation functions manifests a non-perturbative dynamic singularity. This singularity is characterized by a universal power-law exponent that is insensitive to small perturbations.  Explicit realizations of these models in current experiments using polar 
molecules, trapped ions, Rydberg atoms, magnetic atoms, and alkaline-earth and alkali atoms in optical lattices, along with the relative merits and limitations of these different systems, are  discussed.

\end{abstract}



\maketitle

\section{Introduction}

Recent advances in ultracold atom, molecule, and ion experiments~\cite{lamacraft-moore:potential-insights_2012}, 
and the development and application of ultrafast pulsed lasers to probe strongly correlated dynamics in solid-state systems~\cite{gedik:nonequilibrium_2007} have  enabled the experimental study of dynamics in far-from-equilibrium quantum many-body systems~\cite{lamacraft-moore:potential-insights_2012}.  
For example, an abrupt change of parameters in a system's Hamiltonian can create entangled states suitable for quantum metrology and information~\cite{KitagawaUeda_PRA1993,bollinger:optimal_1996,raussendorf:one-way_2001}, enable one to investigate equilibration and thermalization~\cite{cazalilla:focus_2010,polkovnikov:nonequilibrium_2011}, and can be used to characterize fundamental quantum behavior. These experimental capabilities have in turn stimulated a large body of theoretical work, largely because the inherent complexity of far-from-equilibrium interacting quantum systems renders inapplicable most of the standard theoretical tools developed for equilibrium physics.

 In this work we study the time evolution of spin-$1/2$ XXZ spin models following a quantum quench.  We treat systems with varying -- and often arbitrary -- dimensionality and spatial structure of the couplings, including interactions that decay as a power law  with distance, which recently have attracted much attention due to their importance in, for example, ultracold polar molecules~\cite{yan:observation_2013,hazzard:many-body_2014}, Rydberg atoms~
\cite{schwarzkopf:imaging_2011,schwarzkopf:spatial_2013,Schausz:obervation_2012,anderson:dephasing_2002,butscher:atom-molecule_2010,nipper:highly_2012}, magnetic atoms~\cite{lahaye_physics_2009}, and trapped ions~\cite{kim:quantum_2010,britton:engineered_2012}.  
For simplicity we focus most of our attention on initial states that are translationally invariant and
uncorrelated product states, which can easily be created by subjecting an initially spin-polarized system to a strong resonant pulse that initiates the quantum dynamics. The specific initialization protocol  we consider and  the spin models that we study are relevant to numerous ultracold atomic and molecular systems in which the motional degrees of freedom have been frozen out, including trapped ions~\cite{kim:quantum_2010,britton:engineered_2012}, magnetic atoms~\cite{lu:strongly_2011,aikawa:bose-einstein_2012,depaz:resonant_2013}, Rydberg atoms~\cite{schwarzkopf:imaging_2011,schwarzkopf:spatial_2013,Schausz:obervation_2012,anderson:dephasing_2002,butscher:atom-molecule_2010,nipper:highly_2012}, ultracold polar molecules~\cite{yan:observation_2013,hazzard:many-body_2014}, and optical atomic clocks~\cite{swallows:suppression_2011,lemke:p-wave_2011,martin:quantum_2013,Rey_arXiv_2013}, as well as to  condensed-matter systems ranging from nitrogen-vacancy centers in diamond~\cite{doherty:nitrogen-vacancy_2013} and other magnetic defects in solids~\cite{weber:quantum_2010}, to traditional quantum magnets where the spins are realized by electrons localized in a nuclear lattice~\cite{auerbach:interacting_1994,sachdev_quantum_2008,lacroix_introduction_2011,Sachdev_Book_1999}. 
A more comprehensive discussion of  physical realizations is given in Sec.~\ref{app:phys-real}.

\begin{table*}
\setlength{\unitlength}{1.0in}
\includegraphics[width=7in,angle=0]{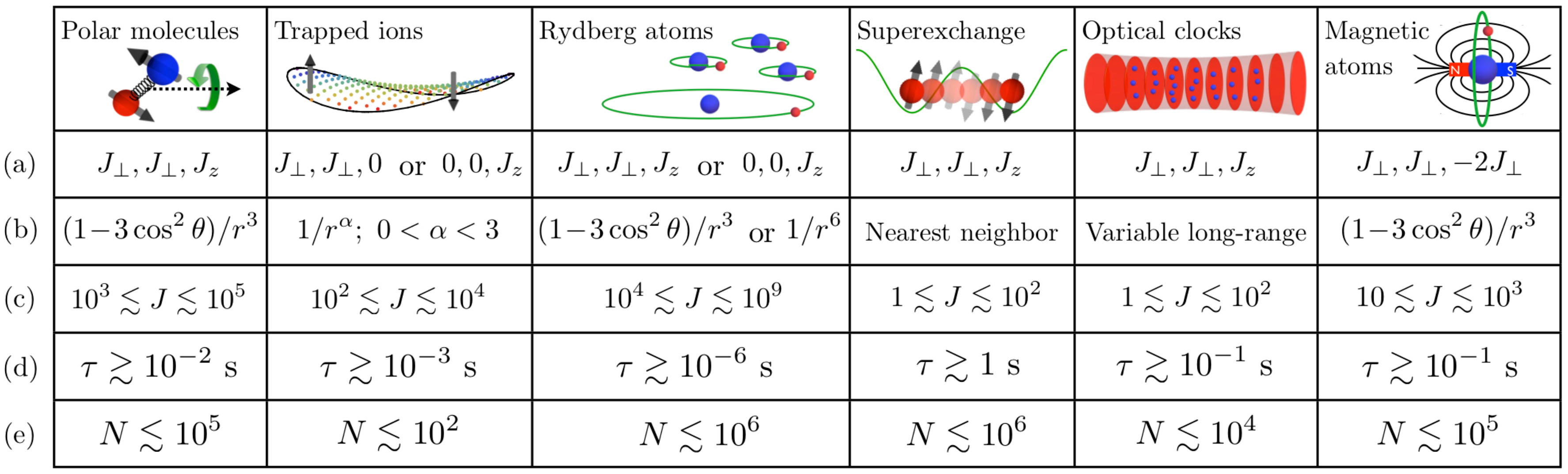}
\caption{
 Properties of several physical systems that can be used to realize the spin models and non-equilibrium dynamics considered in this manuscript. (a) \textbf{Spin-coupling.} Achievable spin couplings, reported as $J_x,J_y,J_z$, are to be understood as coefficients of the Hamiltonian $\mathcal{H}=(1/2)\sum_{i\neq j}\left(J_{x}S_i^xS_j^x+J_{y}S_i^yS_j^y+J_{z}S_i^zS_j^z\right)$. (b) \textbf{Spatial structure.} The distance and angular dependence of the interactions is presented. (c) \textbf{Coupling strengths.} Typical coupling strengths are given as a range (in Hertz), which is intended to reflect reasonable values realizable in current implementations of these systems. (d) \textbf{Coherence times.}
 Coherence times  given are rough lower bounds, but in some special cases (for example by using field-insensitive clock transitions in ions) these times can greatly exceed the stated values. 
  (e) \textbf{Number of spins.} System sizes quoted reflect rough upper limits achieved in current experiments. 
 Rydberg atoms exist in a wide variety of regimes, and the numbers given encompass many different experimental situations.
 \label{tab:systems-overview}}
\end{table*}

The main objective of this work is to characterize and understand the dynamics of local observables, correlations, and entanglement, and their dependence on initial conditions (e.g. spin direction) and range of interactions. Various analytical and numerical techniques are employed, including short-time perturbative methods, exact solutions for the Ising model, and Luttinger liquid theory together with numerically exact methods~\footnote{The ``numerically exact" method employed is the time-dependent density matrix renormalization group (t-DMRG) method~\cite{white:density_1992,vidal:efficient_2004,daley:time-dependent_2004,white:real-time_2004}. What we mean by the term numerically exact is that there is a  parameter that can in principle be increased until the method becomes arbitrarily accurate, and, moreover, that the  errors can be quantified by techniques for scaling in this convergence parameter. Errors are typically expected to be smaller than or comparable to the point size in our plots. } for  one-dimensional spin chains.  We compute the von Neumann entanglement entropy,  
the concurrence, and the  spin squeezing parameter, each of which quantifies a distinct quantum correlation and resource. 
Spin squeezing, for example, quantifies the quantum correlations that are useful for enhanced metrology with sensitivity beyond that achievable with unentangled spins, referred to as the standard quantum limit.

An important issue addressed in this work is the strength of the correlations and of the entanglement obtained in the course of the time evolution.
In general, we find that both correlations and entanglement become significantly larger than they are in the corresponding ground state.  Moreover, at long times, we find evidence that, although not maximal, the entanglement follows a volume law rather than an area law.

An important question regarding the growth of correlations in non-equilibrium systems is whether they exhibit universal behaviors, and if so, how that universality manifests.
We show in a one-dimensional system that universality is realized as a dynamic power-law singularity appearing beyond any order of perturbation theory.
The behavior is controlled by a universal power law exponent; this exponent is universal in the sense that it is insensitive to small perturbations.

We also aim to   provide a brief but fairly comprehensive overview of various experimental systems to which our results apply, highlighting  their unique features and tabulating the form of the interactions, characteristic energy and time scales, and parameter regimes available to each system. Table~\ref{tab:systems-overview} summarizes several of these  and serves as a ``dictionary'' to translate results between systems.

The paper is organized as follows. Section~\ref{sec:models} introduces the XXZ Hamiltonian and the quench  protocol (equivalent to Ramsey spectroscopy) that we study, along with  notation used throughout the rest of the text. Next we calculate dynamics of correlation functions in the short-time limit (Sec.~\ref{sec:short-times}) and the Ising limit (Sec.~\ref{sec:ising-corrn}), and discuss its physical interpretation.  These calculations are primarily a review of prior work, and are included mainly to be self-contained and set the context for calculations of entanglement and universality out of equilibrium  (although they do contain some useful new results and re-arrangements of prior results). Sec.~\ref{sec:ising-entanglement} considers various entanglement measures  for the Ising dynamics, including entanglement entropies for 
two-site ``cut out'' bipartitions, spin squeezing, and concurrence.  Although entanglement is a rather generic feature,  a quantitative analysis comparing the dynamics of these different entanglement measures in a general many-body setting has been lacking. We show that  different types of entanglement, each corresponding to a different quantum resource, manifest at different times  of the Ising dynamics. Sec.~\ref{sec:1D-corrn-entanglement} turns to entanglement in the more general XXZ dynamics in one-dimensional spin chains, where we apply the adaptive time-dependent density matrix renormalization group method (adaptive t-DMRG~\cite{white:density_1992,vidal:efficient_2004,daley:time-dependent_2004,white:real-time_2004}). We quantify the strength of the entanglement emerging in the dynamics by comparing to the entanglement in the corresponding equilibrium systems as well as to maximally entangled states. 
In Sec.~\ref{sec:univ-out-of-eq} we demonstrate an explicit example of universal behavior  out-of-equilibrium. Section~\ref{app:phys-real} summarizes the applicability and relevance of our finding to a variety of atomic, molecular and optical systems, some of which is summarized in Table~\ref{tab:systems-overview}. Finally, Sec.~\ref{sec:conclusions}  concludes and presents an outlook on future theoretical and experimental work.

\section{Hamiltonian and non-equilibrium dynamic procedure \label{sec:models}}


\subsection{Hamiltonian}

We consider the numerous physical systems in atomic  and condensed matter physics that are unified by their description in terms of a spin-$1/2$ XXZ model
\be
H &=& \frac{1}{2}\sum_{i\ne j} \lb J^z_{ij} S^z_i S^z_j
+ \frac{J^\perp_{ij}}{2}\lp S^+_i S^-_j + S^-_i S^+_j \rp\rb, \label{eq:XXZ-model} 
\ee
where the sum extends over all pairs of sites of an arbitrary lattice. By $S_i^z$ and $S_i^\pm=S_i^x\pm i S_i^y$, we denote the usual spin-1/2 operators at site $i$ 
($S_i^{x,y,z}$ are 1/2 times the Pauli operators $\sigma_i^{x,y,z}$). We refer to the first term in Eq.~\eqref{eq:XXZ-model} as the ``direct'' or Ising term, and the second as the ``exchange,'' ``flip-flop,'' or XX term.  The XX terminology comes from rewriting $(1/2)\lp S^+_i S^-_j + S^-_i S^+_j\rp= S^x_i S^x_j+S^y_i S^y_j$, and observing that there are two couplings with the same strength
along the $\hat x$ and $\hat y$ spin directions.   We 
refer to the case when the XX term vanishes as the Ising Hamiltonian and the case when the Ising term vanishes as the XX Hamiltonian. In the specific case where $J^z_{ij}=J^\perp_{ij}$, it is possible to rewrite the Hamiltonian as $H=(1/2)\sum_{i\ne j} J_{ij} \v{S}_i\cdot\v{S}_j$ where $\v{S}_i\equiv (S^x_i,S^y_i,S^z_i)$, which we refer to as the Heisenberg
Hamiltonian.

In our analysis we allow  the  couplings $J^z_{ij}$ and $J^\perp_{ij}$ to take arbitrary values, and unless otherwise specified we do not assume that they exhibit any regularity or translational invariance. Many of our conclusions  thus  apply to disordered systems as well. We also  allow the $(i,j)$-dependence of the couplings to be different for the Ising and XX terms, including distance- and angle-dependence. Similar anisotropies have been considered in models describing orbital magnetism in solid state materials~\cite{kugel:compass_1973,kugel:jahn-teller_1982,nussinov:compass_2013}. They are also important in exactly solvable models harboring topological phases, such as the Kitaev honeycomb model~\cite{kitaev:anyons_2006} and  Yao-Kivelson model~\cite{yao:exact_2007},  and  in models harboring symmetry-protected topological ground states~\cite{liu:symmetry_2012}. Some of these models break the $U(1)$ symmetry assumed here (arising from  the identical strength of the $S^x_i S^x_j$ and $S^y_i S^y_j$ interactions and the absence of cross-terms like $S^x_i 
S^y_j$). However, although the lack of   total  $S^z$   conservation  associated with the broken $U(1)$ symmetry can have  important implications in the dynamics, we nevertheless expect that
several of the features calculated herein (e.g., long distance correlations, entanglement, and non-equilibrium universality) are likely to persist in these  cases as well.  Spin-spin interactions involving three or more sites are also possible, but these  are beyond the scope of the present work.

In many situations, it is useful to add to $H$ single-spin Zeeman terms, given by
\be
H_1 &=& -\sum_i \v{B}_i \cdot \v{S}_i,\label{Beff}
\ee
where the $\v{B}_i$ are local magnetic fields which can vary in space.   
In this paper, such fields are useful for the preparation of initial states for our time evolution, but we only consider dynamics for $\v{B}_i=0$.

To give some idea of the form of the couplings $J^{(\perp,z)}_{ij}$ and the physical systems in which these couplings can arise, we start by  briefly describing five 	disparate physical realizations of Eq.~\eqref{eq:XXZ-model}, chosen to reflect the diversity of these systems: ultracold molecules in optical lattices, trapped ions, Rydberg atoms, ultracold atoms in optical lattices, 
and  ultracold magnetic atoms (see Table~\ref{tab:systems-overview}). 
Section~\ref{app:phys-real} provides a more comprehensive review of realizations of
Eq.~\eqref{eq:XXZ-model}, the physics behind them, and basic characteristics of each system. The following discussion is intended merely to provide context, and is unnecessary for understanding the formal results presented in Secs.~\ref{sec:short-times}-\ref{sec:univ-out-of-eq}.

In {\it ultracold polar molecules} pinned  in optical lattices,  two rotational states can be used to  form the spin-1/2 degree of
freedom, and the spin-spin couplings are induced by dipolar interactions. The difference in dipole moments between the two states (which arises in the presence of an electric field) generates the Ising term, while transition dipole moments between the two rotational  states (which can exist even in the absence of an electric field)   give rise to the spin-exchange terms~\cite{barnett:quantum_2006,wall:hyperfine_2010,wall:emergent_2009,wall:hyperfine_2010,Schachenmayer:dynamical_2010,
 perez-rios:external_2010,herrera:tunable_2010,gorshkov:quantum_2011,gorshkov:tunable_2011}. Unlike the nearest-neighbor interactions arising from superexchange, the dipolar interactions are long-ranged and anisotropic. For the choices of rotational states used so far in experiments \cite{yan:observation_2013,hazzard:many-body_2014} $J^z_{ij}$ and $J^\perp_{ij}$ are  both proportional to $(1-3\cos^2\Theta_{ij})/r_{ij}^3$ where $r_{ij}$ is the distance between the dipoles and $\Theta_{ij}$ is the angle between the inter-molecular axis and the quantization axis provided by the external  field (electric or magnetic). More complicated spin-spin interaction anisotropies, which can even break the $U(1)$ symmetry, can be  generated by more general choices of rotational states and/or by microwave dressing \cite{gorshkov:quantum_2011,gorshkov:tunable_2011,wall:hyperfine_2010,manmana:topological_2013,gorshkov:kitaev_2013}.
 Other implementations of spin models in polar molecules using hyperfine levels to encode the spin have also been proposed, for example see  Ref.~\cite{carr:cold_2009} and references therein.

In one- and two-dimensional {\it crystals of trapped ions}, 
hyperfine states can realize a spin-1/2. By addressing the ions with a spin-dependent optical potential, the vibrations of the crystal mediate a long-range Ising interaction that can be approximately described by a spatial power law $J_{ij}\propto 1/r_{ij}^\alpha$, with $0\le \alpha < 3$, where $r_{ij}$ is the distance between ions $i$ and $j$~\cite{porras:quantum_2006,kim:entanglement_2009,kim:quantum_2010,barreiro:open-system_2011, britton:engineered_2012,islam:emergence_2013}. To engineer an XX model, it suffices to add a strong transverse  field that projects out the off-resonant terms  in the Ising interactions that change the  magnetization
along the field quantization direction~\cite{Richerme_arXiv_2014}. More general XXZ models  can be implemented,  for example, by using multiple spin-dependent optical potentials~\cite{porras:effective_2004}.

In {\it frozen Rydberg gases} the Ising-type Hamiltonian can be realized via the strong Rydberg-Rydberg van der Waals interaction~\cite{weimer:quantum_2008}. Typical experiments are very fast and limited by the Rydberg lifetime, which in turn guarantees that motional degrees of freedom remain frozen even without an underlying lattice potential. To obtain an (anisotropic) $1/r^3$ first-order dipolar potential between resonant dipoles oscillating between neighboring Rydberg states, one can employ so-called F{\"o}rster resonances.   As only the size of the Rydberg atoms limits the dipole moment, the resulting interaction strength can be very large and scales in the dipolar case  as $n^4$ with the principal quantum number $n$.

For {\it ultracold spin-1/2  atoms}  loaded into optical lattices, spin models emerge in Mott-insulating states, in which on-site interactions much stronger than the tunneling pin the lattice filling (for a range of chemical potentials) to integer numbers of atoms per site. In these systems, two hyperfine states encode the spin-1/2 degree of freedom  and  superexchange processes~\cite{trotzky:time-resolved_2008} lead to spin-spin interactions of the XXZ form in Eq.~\eqref{eq:XXZ-model}. In this implementation  the  $J_{ij}$ couplings are restricted to  nearest neighbors. For spin-independent lattices, fermions realize a Heisenberg model, while for bosons the spin model is  XXZ and depends on the relative sizes of the three scattering lengths between the two spin states, $a_{\uparrow\uparrow}$, $a_{\downarrow\downarrow}$, and $a_{\uparrow\downarrow}$. More general XXZ models can be realized with spin-dependent lattices.

Spin-exchange interactions also occur in {\it magnetic atoms}, which therefore realize spin models when confined in optical  lattices~\cite{hensler:dipolar_2003}. In this case the spin degree of freedom is encoded in atomic hyperfine states (and in general it is not restricted to be spin $1/2$). Although in general magnetic dipole interactions include magnetization-changing terms~\cite{hensler:dipolar_2003} not accounted for in an XXZ spin model,  these terms can be  energetically  suppressed at   high enough magnetic field.  The magnetization-conserving  dipolar interactions remain resonant, and implement an XXZ spin model. In recent experiments carried out with Cr atoms with effective spin $S=3$, the corresponding spin Hamiltonian realizes $J^z=-2J^\perp$~\cite{depaz:nonequilibrium_2013}.

\subsection{Dynamic procedure}
\begin{figure}
\setlength{\unitlength}{1.0in}
\includegraphics[width=3.4in,angle=0]{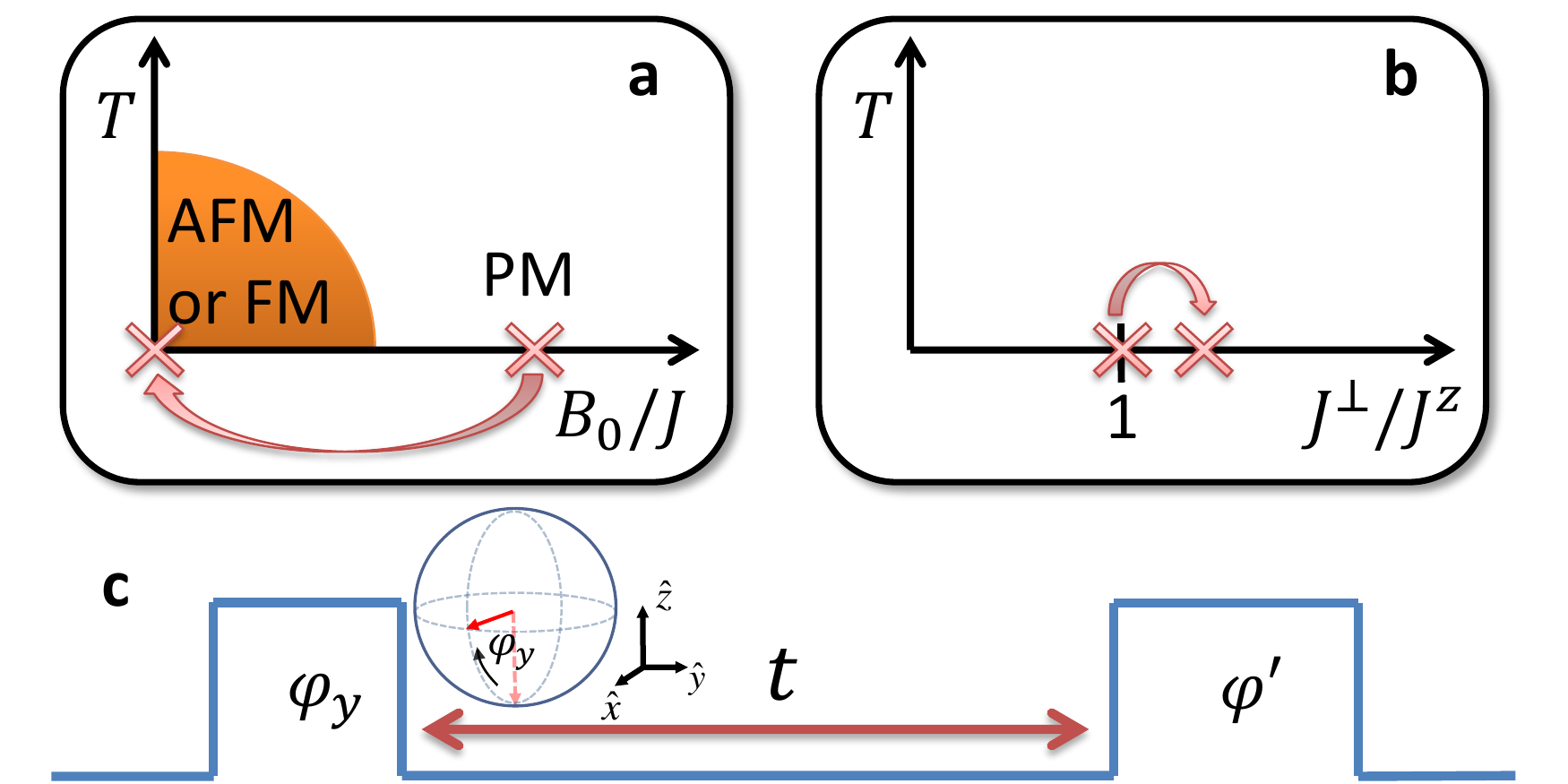}
\caption{ 
Three interpretations of the dynamic protocol that we study. (a) Sudden quench from the XXZ Hamiltonian with a strong magnetic field $B_0 {\hat n}$ polarizing the spins in the $\hat n$ direction to zero field $B=0$ for times $t>0$, followed by a measurement of the observables at time $t$ after the quench. (b) 
Sudden quench from the XXZ Hamiltonian at a ferromagnetic SU(2) symmetric point ($J_{ij}^z=J_{ij}^\perp$), where the ground states consist of all spins polarized along the same axis, to the XXZ Hamiltonian of interest.  One adds an otherwise inconsequential infinitesimal magnetic field to choose the proper initial ground state. (In fact this interpretation holds for antiferromagnetic interactions as well, since the dynamics is independent of multiplying the Hamiltonian by $-1$.)
 (c) Ramsey spectroscopy: initial spins along the ``down'' ($-\hat z$) spin direction are rotated by an angle $\varphi$ about the $\hat y$ spin direction. These evolve under the XXZ  Hamiltonian \eqref{eq:XXZ-model} for an evolution time $t$. An arbitrary component of the Bloch vector can be read out by proper choice of the area ($\varphi'$) and phase of the second pulse, and other correlation functions can be measured by high-resolution imaging and/or spin manipulation.
 \label{fig:dynamics-cartoon}}
\end{figure}

Figure~\ref{fig:dynamics-cartoon} depicts the dynamic procedure considered in this manuscript.
All spins are initially aligned (at time $t=0$) along some direction $\hat n$, and then this
initial state $\ket{\Psi(0)}=\bigotimes_i \ket{\hat n}_i$ evolves under the XXZ Hamiltonian
\eqref{eq:XXZ-model} for a time $t$. Then one measures time-dependent expectation values of observables $\langle
\mathcal{O}(t) \rangle$, where the expectation value is taken in the time-evolved
 state $e^{-iHt}\ket{\Psi(0)}$.
Because the interaction Hamiltonian possesses a U(1) symmetry associated with rotational invariance about the spin $z$-axis, we can take the initial vector $\hat n$ to lie in the $x$-$z$ plane, $\hat n = (\sin\varphi,0,-\cos\varphi)$, without loss of generality. We will refer to this angle $\varphi$ as the ``tipping angle.''

This protocol has been used to observe dipolar spin-exchange interactions in ultracold molecules~\cite{yan:observation_2013,hazzard:many-body_2014}, to benchmark quantum simulators of hundreds of trapped ions~\cite{britton:engineered_2012}, and to precisely measure atomic transitions as well as many-body interactions  in optical lattice clocks~\cite{hinkley:atomic_2013,bloom:optical_2014,martin:quantum_2013,lemke:p-wave_2011}.
Related protocols have been used or proposed  for measuring spin relaxation, diffusion, and
transport~\cite{lewandowski:observation_2002,mcguirk:normal-superfluid_2003,kuklov:detecting_2004,du:observation_2008,natu:anomalous_2009,piechon:cumulative_2009,
mischuck:coherent_2010,maineult:spin_2012,pechkis:spinor_2013,
koschorreck:universal_2013},
for determining many-body interactions~\cite{anderson:dephasing_2002,widera:quantum_2008,
gibble:decoherence_2009,rey:many-body_2009,butscher:atom-molecule_2010,yu:clock_2010,nipper:highly_2012,
britton:engineered_2012,
hazlett:anomalous_2013,martin:quantum_2013,
yan:observation_2013},
for probing real-space correlations~\cite{knap:probing_2013,kitagawa:ramsey_2010},
and for characterizing topological order~\cite{abanin:interferometric_2013,atala:direct_2013}.

Despite the simplicity of this protocol, we will see that the dynamics develops strong, long-ranged correlations, entanglement, and interesting dynamics that is frequently not captured by mean-field theory. Many of our results generalize to the case where the initial direction $\hat n$ can vary for each spin (some generalizations have been discussed in~\cite{FossFeig_PRA2013,FossFeig_NJP2013,vdW_NJP2013}) but here for simplicity we restrict our attention to uniform $\hat n$.

The dynamic protocol can be  viewed either as Ramsey spectroscopy (illustrated in Fig.~\ref{fig:dynamics-cartoon}c), or as one of two different types of quantum quenches.  From the perspective of Ramsey spectroscopy, the protocol begins by preparing all particles in the same internal state, which we will consider to be the ``spin-down'' state.  This preparation can typically be achieved with very high fidelity in ultracold atomic and molecular system, e.g. via optical pumping.  An effective transverse field $B \sum_i S^y_i$ is then pulsed on for a time $\tau$ chosen to rotate each spin to the desired state $\hat n$ in the $x$-$z$ plane by an angle $\varphi=B\tau$.  This rotation requires that $B\gg \{J^z_{ij},J^\perp_{ij}\}$ (and thus, for a fixed $\varphi$, that $\{J^z_{ij}\tau,J^\perp_{ij}\tau\}\ll 1$), so that we may neglect interactions during the pulse.  After evolving for a time $t$ under the interaction Hamiltonian, a final pulse rotates a desired Bloch-vector component onto the $z$ axis, where it can 
be measured.  The experimental appeal of the dynamic protocol considered in this paper is now evident: While cooling to ground states of interacting spin Hamiltonians can be extremely challenging in atomic and molecular systems, the preparation of each individual spin in its ``down'' state and the application of a pulsed effective transverse magnetic field is straightforward.

From the perspective of a quantum quench, the system starts in the ground state of an initial Hamiltonian
(discussed below), the Hamiltonian is then suddenly switched to the XXZ Hamiltonian in
Eq.~\eqref{eq:XXZ-model}, and the state is allowed to evolve for a time $t$ before being measured.  The first
quench interpretation (illustrated in Fig.~\ref{fig:dynamics-cartoon}a) follows by considering the initial
state to be the ground state of the Hamiltonian $H+H_{1}$, with $\v{B}_i=B_0 \hat n$, in a large ($B_0\gg
J^z,J^{\perp}$) magnetic field along the $\hat n$ direction,  i.e. the ground state of $H_1$.  The dynamics is
then induced by suddenly turning off the magnetic field (from $H_1$) at time $t=0$, leaving only the XXZ
Hamiltonian $H$ in Eq.~\eqref{eq:XXZ-model}.  The second quench interpretation (illustrated in
Fig.~\ref{fig:dynamics-cartoon}b) is to consider the initial state as the ground state of an initial
Heisenberg Hamiltonian:  $J^z_{ij}=J^\perp_{ij}<0$~\footnote{The state includes no fluctuations from this;
formally this is easiest to see by considering the state $\ket{\Psi(0)}\equiv\langle\cdots \downarrow
\downarrow\cdots\rangle$ [note that since the Hamiltonian for $J^z_{ij}=J^\perp_{ij}$ is SU(2) symmetric for
spin rotations, all rotations of this state are degenerate]. When $H$ acts on this state, the flip-flop terms
vanish, and the action of the Ising term is a c-number times the state; this confirms that $\ket{\Psi(0)}$ is
an eigenstate. Using $\langle\v{S}_i\cdot \v{S}_j\rangle\le1/4$, it can be shown that it saturates the minimum
possible energy for the Hamiltonian (ground state) if $J^\perp_{ij}<0$.}. The states can be massively
degenerate for large systems, corresponding to different directions in which the spins align. To break this potential degeneracy and choose the
desired initial state with all spins along $\hat n$, we can add a field $\v{B}_i=\epsilon\hat n$ for an
infinitesimal $\epsilon=0^+$. In this way we can view the quench as being from the ground state of an initial
SU(2) invariant Hamiltonian with $J^z_{ij}=J^\perp_{ij}$ to the more general XXZ Hamiltonian in
Eq.~\eqref{eq:XXZ-model}.  This perspective is particularly useful when Eq.~\eqref{eq:XXZ-model} has
$J^z_{ij}\approx J^\perp_{ij}$, such that the quench can be viewed as a small change in the Hamiltonian
parameters. This is relevant to the emergence of universality out-of-equilibrium studied in
Sec.~\ref{sec:univ-out-of-eq}.  The sign of the overall Hamiltonian is irrelevant for the dynamics, so the
quench can be regarded as a perturbation  if $J^z_{ij}\approx J^\perp_{ij}$, regardless of the overall sign of
the couplings.

Before turning to our results, we emphasize that the Hamiltonian \eqref{eq:XXZ-model} is time-independent. (Although the Ramsey spectroscopy's preparation and readout may be viewed as a time-dependent $H_1$ applied before and after the dynamics of interest.) 
Further interesting features may arise by considering dynamic coefficients, but they are beyond the scope of this paper.

\section{Short-time results\label{sec:short-times}}

For short times satisfying $\{J^\perp_{ij} t,J^z_{ij} t\} \ll 1$, one can calculate correlation functions using  time-dependent perturbation theory. The expectation value of an operator $\mc O$ evolves in time under a time-independent Hamiltonian $H$ as $\expec{\mc O(t)}= \expec{\mc O}-it\expec{[\mc O,H]}-\frac{t^2}{2} \expec{[[\mc O,H],H]} + O(t^3)$, where the expectation value is taken in the initial state. We can evaluate these commutators and expectation values for the dynamics of interest, at least for low orders of perturbation theory.

Defining
\begin{equation}
{\mc C}^{ab}_{ij} \equiv \expec{S^a_iS^b_j}
\end{equation}
with $a,b \in \{x,y,z,+,-\}$, for $i\ne j$ we find
\begin{eqnarray} \expec{S_i^x} &=& \frac{1}{2}\sin \varphi \left\{1-\frac{t^2}{8}\lb \Xi^{(2)}_i \sin^2\varphi +(\Xi^{(1)}_i)^2\cos^2\varphi \rb\right\} \nonumber \\
&&\hspace{1.4in}{}
+O(t^4),
\nonumber\\
\expec{S_i^y} &=& - \frac{1}{2}\sin \varphi \left\{ \frac{t}{2} \Xi^{(1)}_i\cos(\varphi) \right\}+O(t^3) ,
\nonumber\\
{\mc C}^{xy}_{ij} &=& \frac{t \sin(2\varphi)\sin\varphi }{16} \lb (J^z_{ij}-J^\perp_{ij})-\Xi^{(1)}_j \rb + O(t^2),  \label{eq:short-time-corrns}
\nonumber
\\
{\mc C}^{yz}_{ij} &=& \frac{t}{8}\lb\frac{ \sin(2\varphi)\cos\varphi }{2} \Xi^{(1)}_i
    + (J^z_{ij}-J^\perp_{ij} ) \sin^3\varphi \rb \nonumber \\
    &&\hspace{1.4in}{}+ O(t^2),\label{eq:short-time-formulas}
\end{eqnarray}
where  
we have defined
\be
\Xi_i^{(m)} &=& \sum_{j \ne i} \lp J_{ij}^z-J_{ij}^\perp \rp^m.
\ee 
The total $z$ axis magnetization is conserved for all times: $\sum_i\expec{S^z_i}(t)=-(N/2)\cos\varphi$ for $N$ spins. 
The correlators not shown (and those not trivially related to the shown correlators above by transposing the indices) are given by their $t=0$ values to linear order.
Interestingly, the results given in Eq.~\eqref{eq:short-time-formulas} depend only on $J^\perp-J^z$, but this is \textit{not} expected to generalize to higher orders in time; already at quadratic order, terms such as $[(J^\perp)^2-(J^z)^2]t^2$ can occur in addition to $(J^\perp-J^z)^2t^2$ terms.
The results for the case $J^\perp_{ij}=\lambda J^z_{ij}$ and assuming translational invariance were given in Ref.~\cite{Hazzard}, which focused on dipolar interactions $J^z_{ij}\propto (1-3 \cos^2\Theta_{ij})/r_{ij}^3$, with $\Theta_{ij}$ the angle between a quantization axis and the interparticle separation. The more general results given by Eq.~\eqref{eq:short-time-formulas}  are useful for finite or disordered systems, and for compass-type models where the spatial anisotropy and position-dependence may be different for the Ising and spin-exchange terms~\cite{kugel:compass_1973,kugel:jahn-teller_1982,nussinov:compass_2013,kitaev:anyons_2006,liu:symmetry_2012,
manmana:topological_2013,gorshkov:kitaev_2013}.

The connected correlation functions
\begin{equation}
{\mc G}^{ab}_{ij} \equiv \expec{S^a_iS^b_j}-\expec{S^a_i}\expec{S^b_j} 
\label{eq:correlators-defn}
\end{equation}
immediately follow from the results in Eq.~\eqref{eq:short-time-formulas},  
\be
{\mc G}^{xy}_{ij} &=& \frac{t \sin(2\varphi)\sin\varphi }{16} \lp J^z_{ij}-J^\perp_{ij}\rp + O(t^2),
\nonumber
\\
{\mc G}^{yz}_{ij} &=& \frac{t  \sin^3\varphi }{8}\lp J^z_{ij}-J^\perp_{ij} \rp + O(t^2).
\label{eq:short-time-connected-formulas}
\ee

The single-spin expectation values and correlation functions calculated here are directly accessible in experiments. For example ${\mc C}_{ij}^{ab}$, which will be crucial input for the entanglement measures considered below, can be measured by high-resolution imaging or related to scattering results.
These short-time expressions provide insight into the many-body dynamics. For example, one can see two distinct contributions to $\expec{S_i^x}$: a decrease due to precession, proportional to $\cos^2\varphi $, and a decrease associated with a shrinking Bloch vector length of that spin due to entanglement with the other spins, proportional to $\sin^2\varphi $. The former can be understood within mean-field theory, while the latter is due to quantum fluctuations and cannot. The  tipping angle dependence also  suggests the parameter regime in which a mean-field description of the Bloch vector dynamics is valid:   for small tipping angles, $0<\varphi\ll \pi/2$, mean-field effects dominate the dynamics, while as $\varphi$ approaches $\pi/2$ the dynamics  is dominated by  genuine many-body correlations. The short-time formulas are also useful for estimating the convergence of new approximate 
theories, such as the ``moving average cluster expansion'' developed in Ref.~\cite{hazzard:many-body_2014}.

\section{Ising limit correlations and entanglement \label{sec:ising-corrn}}

\subsection{Review of correlations in the Ising limit \label{sec:ising-correlation}}

Here we give an overview of the Bloch vector  and correlation function dynamics in the Ising limit ($J_\perp=0$). These observables are the basis  for the  computation of the entanglement entropy discussed  in Sec.~\ref{sec:ising-entanglement}. We will also make use of the Ising correlation functions  in Sec.~\ref{sec:1D-corrn-entanglement} when comparing them to numerical results for the XXZ model in one dimension.  There, we discuss which features are particular to the Ising model and which generalize to the XXZ model.

While the equilibrium properties of the Ising model are classical, this is not the case for its dynamical properties, and  mean-field theory  fails to capture the dynamics in many regimes. For example, for an initial $\varphi=\pi/2$ tipping angle, mean-field theory predicts that there is no dynamics, while in fact evolution occurs on a timescale set by $J^z$. Furthermore, starting from a product state the dynamics generates entangled states, including cluster states that suffice for one-way quantum computation~\cite{raussendorf:one-way_2001}, spin squeezed states enabling quantum metrology~\cite{KitagawaUeda_PRA1993,Hazzard}, and Greenberger-Horne-Zeilinger (GHZ) states~\cite{bollinger:optimal_1996,greenberger:bells_1989} --- a type of Schr{\"o}dinger cat state.  Nevertheless, the Ising limit does possess a special structure that facilitates the exact calculation of arbitrary-order correlation functions~\cite{FossFeig_PRA2013,FossFeig_NJP2013,vdW_NJP2013}, and leads to unique dynamical features.  In this work, we focus on the  single-spin and two-spin expectation values, presented below.

In the Ising limit $J_\perp=0$, the XXZ Hamiltonian \eqref{eq:XXZ-model} reduces to
\begin{equation}
 H_{\text{Ising}} = \frac{1}{2}\sum_{i\ne j} J^z_{ij} S^z_i S^z_j.
 \label{eq:Ising-model}
\end{equation}
Because all of the terms in this sum commute with one another, the time evolution operator
$U=e^{-iH_{\text{Ising}}t}$ factorizes, rendering analytic calculations possible
\cite{Emch66,Radin70,Kastner11,Kastner12}. Exact results have been derived
for correlation functions of the Ising Hamiltonian \eqref{eq:Ising-model} with arbitrary pure product initial
states, even  in the presence of  general Markovian decoherence mechanisms~\cite{FossFeig_NJP2013,
FossFeig_PRA2013}, and for the coherent evolution of initial mixed product  states \cite{vdW_NJP2013}.
Here we limit our calculations to  initial states of the form
\begin{equation}
 |\Psi(0)\rangle = \bigotimes_j \bigg( e^{i \phi_j/2} \cos\frac{\theta_j}{2}
| \uparrow \rangle_j+ e^{-i \phi_j/2}\sin\frac{\theta_j}{2}| \downarrow
\rangle_j \bigg), \label{eq:initialstate}
\end{equation}
where $\theta_j$ and $\phi_j$ are the polar and azimuthal angles in the Bloch sphere representation, and $| \uparrow \rangle_j $ and $| \downarrow \rangle_j$ are eigenstates of $S^z_j$ with eigenvalues $+1/2$ and $-1/2$, respectively.

Since $S^z_j$ commutes with $H_{\rm Ising}$, its expectation value remains constant during the time evolution,
\be
\langle S^z_j \rangle(t) & =&\frac{1}{2}\cos\theta_j,
\nonumber \\
\mathcal{C}^{zz}_{jk}(t) &=& \frac{1}{4}\cos\theta_j \cos\theta_k.
\ee
The expectation value of $S^{\pm}_j = S^x_j \pm i S^y_j$
is
\begin{equation}
 \langle S^+_j \rangle(t)=\frac{1}{2}e^{i\phi_j} \sin\theta_j \prod_{k\neq j} g^{+}_k(J_{kj}t),
 \label{eq:ising-Splus-expectation}
\end{equation}
where
\begin{equation}
 g^{\pm}_{j}(x) \equiv \cos^2(\theta_j/2)e^{-ix/2} \pm \sin^2(\theta_j/2)e^{ix/2},
\end{equation}
and the $x$ and $y$ spin components are then given by
\begin{align}
 \langle S^{x}_j \rangle = \text{Re} \langle S^+_j \rangle ,\quad
 \langle S^{y}_j \rangle = \text{Im} \langle S^+_j \rangle .
\end{align}
The remaining  two-point correlation functions are given by
\begin{align}
\label{eq:Cpz}
 \mathcal{C}^{+z}_{jk}(t) &= \frac{1}{4} e^{i\phi_j} \sin\theta_j g^{-}_{k}(J_{jk}t)
 \prod_{l\neq j,k} g^{+}_{l}(J_{jl}t),\\
\label{eq:Cpp}
 \mathcal{C}^{++}_{jk}(t) &= \frac{1}{4} e^{i(\phi_j + \phi_k)} \sin\theta_j \sin\theta_k
 \prod_{l\neq j,k} g^{+}_{l}(J_{jl}t + J_{kl}t),\\
\label{eq:Cpm}
 \mathcal{C}^{+-}_{jk}(t) &= \frac{1}{4} e^{i(\phi_j - \phi_k)} \sin\theta_j \sin\theta_k
 \prod_{l\neq j,k} g^{+}_{l}(J_{jl}t - J_{kl}t),
\end{align}
From the definitions, it follows that
 $\mathcal{C}^{+-}_{jk}(t) = \mathcal{C}^{-+}_{jk}(t)^*$,
 $\mathcal{C}^{++}_{jk}(t) = \mathcal{C}^{--}_{jk}(t)^*$,
and $\mathcal{C}^{+z}_{jk}(t) = \mathcal{C}^{-z}_{jk}(t)^*$,
and the $x$ and $y$ components of the spin correlators are given by 
\begin{align}
 \mathcal{C}^{x x}_{ij} &=
\frac{1}{4}\left(\mathcal{C}^{++}_{ij} + \mathcal{C}^{--}_{ij}+\mathcal{C}^{+-}_{ij} + \mathcal{C}^{-+}_{ij} \right)\!,
\\
\mathcal{C}^{y y}_{ij} &=
 \frac{1}{4}\left(\mathcal{C}^{+-}_{ij} + \mathcal{C}^{-+}_{ij}-\mathcal{C}^{++}_{ij} - \mathcal{C}^{--}_{ij} \right)\!,
 \\
 \mathcal{C}^{x y}_{ij} &=
 \frac{1}{4i}\left(\mathcal{C}^{++}_{ij} - \mathcal{C}^{--}_{ij}-\mathcal{C}^{+-}_{ij} +\mathcal{C}^{-+}_{ij} \right)\!.
\end{align}
The connected correlators ${\mc G}^{ab}_{ij}$ follow  from the expressions above. One finds
\begin{align}
{\mc G}^{+z}_{jk} &= \frac{e^{i\phi_j}}{4} \sin\theta_j  \prod_{l\ne j,k} g_l^+(J_{jl} t)
\nonumber \\
 &\hspace{0.5in}\times 
 \lb g_k^-(J_{jk}t)-\cos \theta_j g_k^+(J_{jk} t) \rb\!, \\
{\mc G}^{++}_{jk} &= \frac{e^{i\lp \phi_j+\phi_k \rp}}{4} \sin\theta_j\sin\theta_k\bigg[ \prod_{l\ne j,k} g_l^+(J_{jl}t+J_{lk}t) \nonumber \\
		&- g_k^+(J_{jk}t) g_j^+(J_{jk}t) \prod_{l \ne j,k} g_l^+(J_{jl}t) g_l^+(J_{l k}t) \bigg]\!, \\
{\mc G}^{+-}_{jk} &= \frac{e^{i\lp \phi_j-\phi_k \rp}}{4} \sin\theta_j\sin\theta_k\bigg[ \prod_{l\ne j,k} g_l^+(J_{jl}t-J_{lk}t) \nonumber \\
		&- g_k^+(J_{jk}t) g_j^+(-J_{jk}t)\prod_{l \ne j,k} g_l^+(J_{jl}t) g_l^+(-J_{l k}t) \bigg]\! \label{eq:ising-connected-corrn}.
\end{align}

All higher-order correlation functions of $S^\pm$ and $S^z$ involving more than two spins, such as $\expec{S^a_iS^b_jS^c_k}$ with $\{a,b,c\}\in \{\pm ,z\}$, have simple exact expressions as well (see for example  Refs.~\cite{FossFeig_PRA2013,FossFeig_NJP2013,vdW_NJP2013}), but we omit them here because they are unnecessary for our purposes.

\subsection{Entanglement in the Ising limit \label{sec:ising-entanglement}}

\begin{figure*}
\setlength{\unitlength}{1.0in}
\includegraphics[width=2\columnwidth,angle=0]{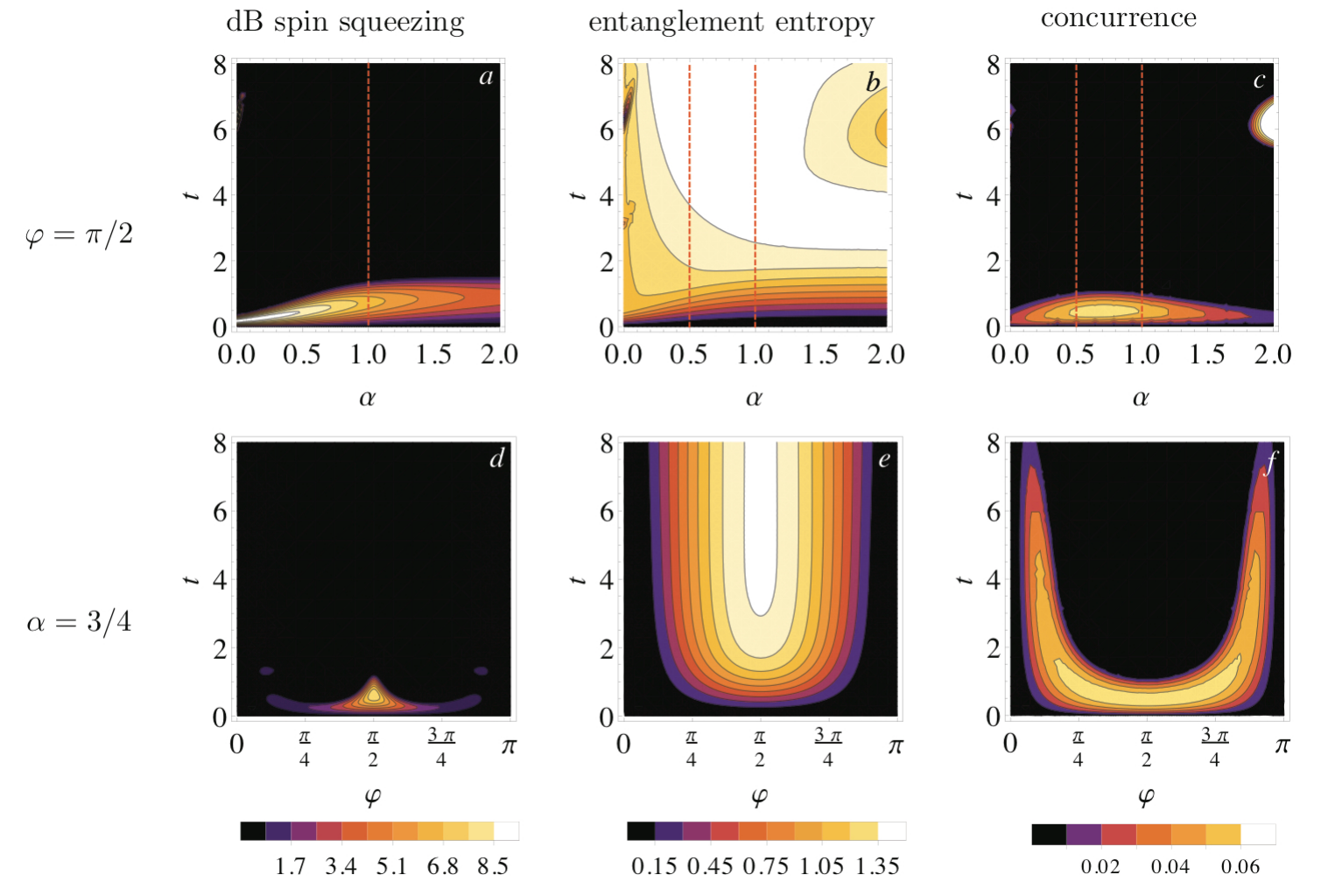}
 \caption{
  Time evolution of three entanglement measures    
  for  Ising chains with $L=41$ lattice sites.
Top row: dB spin squeezing (left), entanglement entropy $S_{\mc R_{ij}}$ (middle), and concurrence (right) 
in the $(\alpha,t)$-plane, starting from a fully $x$-polarized $\varphi=\pi/2$ initial state.
Bottom row: As above, but in the $(\varphi,t)$-plane for an interaction exponent $\alpha=3/4$. The region ${\mc
R_{ij}}$ is chosen to consist of next-nearest neighbor spins at the center of the system. The same qualitative features
are seen for alternative  lattice geometries and for values of $i$ and $j$ that are not necessarily adjacent and at the center of the system. Vertical, dashed  red lines indicate $\alpha=d$ and $\alpha=d/2$, values where the character of the dynamics changes as discussed in the text.
 \label{fig:ising-entanglement}}
\end{figure*}

In this section we calculate and compare the dynamics of distinct entanglement measures   in the Ising limit.   
Each measure quantifies a distinct quantum correlation and resource~\cite{horodecki:quantum_2009}, and  they 
help to classify and understand the structure of many-body phases of matter and dynamics~\cite{cirac:entanglement_2012}.

An example of the qualitatively different dynamics of these distinct entanglement measures is furnished by the well understood all-to-all coupling limit, $J_{ij}=J$ for all $i$ and $j$, where Eq.~\eqref{eq:Ising-model} is known as the one-axis twisting Hamiltonian \cite{KitagawaUeda_PRA1993}. For an initial tipping angle $\varphi=\pi/2$, spin squeezing (one type of entanglement) emerges in this model at short times \cite{KitagawaUeda_PRA1993}, while at later times an entangled   GHZ state is generated (more discussion and definitions are given below)\cite{molmer:multiparticle_1999}. A spin-squeezed state   exhibits enhanced phase sensitivity with respect  to the shot noise limit in standard Ramsey spectroscopy measurements but under Ising dynamics never reaches the Heisenberg limit (the maximum sensitivity allowed by quantum mechanics). In contrast, a GHZ state can reach the Heisenberg limit but, utilizing it in spectroscopy requires a modified Ramsey sequence, with  a final readout based, for  example, on  spin parity measurements $(-1)^{2S^z}$ \cite{bollinger:optimal_1996}.

Away from the all-to-all limit, the dynamics is much less well understood; calculating and understanding it are the main focuses of this section. For illustrative purposes, the coupling constants  in Eq.~\eqref{eq:XXZ-model} are chosen to be $J^z_{ij}=J/r_{ij}^{\alpha}$  where $r_{ij}$ denotes the distance between spins $i$ and $j$. For simplicity we consider one-dimensional lattices (i.e. chains), but our qualitative conclusions do not depend on this. We study the dependence of different entanglement measures on the initial tipping angle ($\phi_j=0$ and $\theta_j=-\varphi$)  and the range of interactions ($\alpha$).

\textit{Spin squeezing.}---Spin squeezing, which was introduced for the first time in Ref.~\cite{KitagawaUeda_PRA1993}, characterizes the sensitivity of a
state to SU(2) rotations, and is relevant for both entanglement detection (it is an entanglement witness)  and quantum metrology (see  Refs.~\cite{PhysRevA.46.R6797,Ma2011} for more complete reviews).
It also provides a lower bound on the minimum size of
a genuine many-body entangled subsystem~\cite{PhysRevLett.86.4431}.
Spin squeezed states can be visualized as states
with anisotropic fluctuations of the spin vector in the directions perpendicular to the mean spin. Roughly speaking, a  quantum state is considered spin squeezed if the variance of one spin component is smaller than that  of an  uncorrelated spin coherent state. Due to the Heisenberg uncertainty relation, reduction of the variance in one direction causes an increase of fluctuations in the other.
Spin squeezing is relatively easy to  visualize, generate, and measure experimentally since it only involves the  first and second moments of the collective angular momentum operators.

There are multiple definitions of spin squeezing, depending on the context where it is used. Here we adopt the squeezing parameter introduced  by  Wineland {\it et al.} in the context of Ramsey spectroscopy \cite{PhysRevA.46.R6797}
\be
 \xi = \min_{{\hat n}}\frac{\sqrt{N} \sqrt{\langle \left(\v{S} \cdot \hat{n}
\right)^2 \rangle - \langle \v{S} \cdot \hat{n} \rangle ^2} }{ |\langle
\v{S} \rangle|},
\ee
where the minimization is over unit vectors $\hat n$  perpendicular to the mean spin direction $\expec{\sum_i \v{S}_i}$.  The correlation functions in the definition of $\xi$ are readily evaluated using Eqs.~\eqref{eq:Cpz}-\eqref{eq:Cpm} of the previous section.  A state is then said to be squeezed when $\xi<1$.  It is common to report
squeezing in decibels, 
\be
 \text{dB squeezing} = -10 \log_{10} \xi^2,
\ee
such that a state is squeezed when dB squeezing is positive.  We note that squeezing is an entanglement witness for general mixed states: when dB squeezing is positive, the state is necessarily entangled \cite{Sorensen2001,Korbicz_PRL2005,Ma2011}.

In Ref.~\cite{KitagawaUeda_PRA1993} the generation of spin squeezing was discussed for the case of a  one-axis twisting (OAT) Hamiltonian $H_{\text{OAT}}= J\left(S^z\right)^2/2$, corresponding to the $\alpha=0$ case of the Ising Hamiltonian Eq.~\eqref{eq:Ising-model}. The OAT Hamiltonian generates spin-squeezed states at short times, a fact that has been confirmed in a number of experiments \cite{Gross_Nature2010,Esteve_Nature2008}, but the squeezing is transient and  disappears at longer times. However, the loss of spin squeezing does not necessarily imply a loss of entanglement --- for example, at longer times ($t=\hbar\pi/J$) entangled GHZ-states occur \cite{molmer:multiparticle_1999}.

\begin{figure}
\setlength{\unitlength}{1.0in}
 \includegraphics[width=\linewidth]{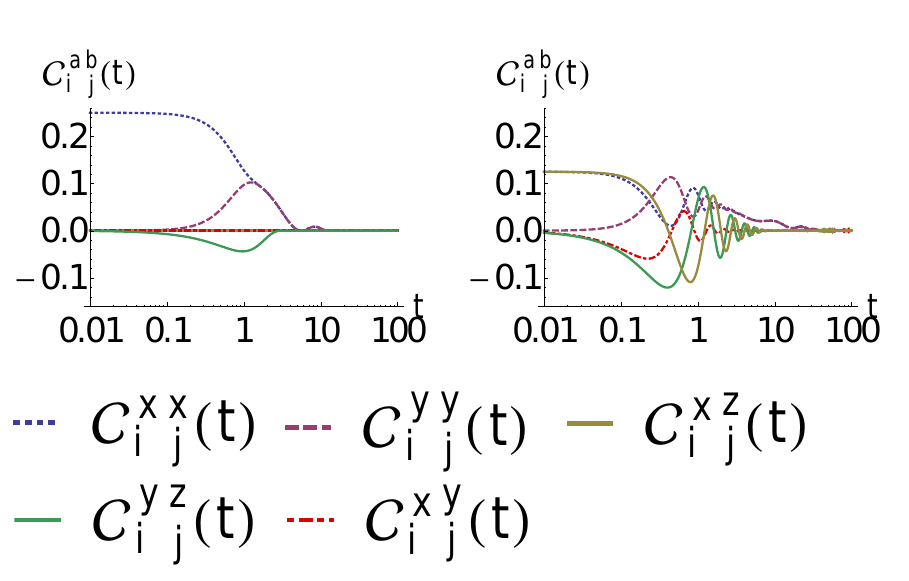}
 \caption{Time evolution of the two-spin correlation functions
$\mathcal{C}^{xx}_{ij}(t)$,
$\mathcal{C}^{yy}_{ij}(t)$, $\mathcal{C}^{xy}_{ij}(t)$, $\mathcal{C}^{xz}_{ij}(t)$ and
$\mathcal{C}^{yz}_{ij}(t)$ for a
power law interacting Ising chain consisting of 41 lattice sites. The interaction exponent is $\alpha=3/4$,
with initial tipping angles $\varphi=\pi/2$ (left) and $\varphi=\pi/4$ (right). Lattice sites $i$ and $j$ are
chosen one lattice spacing to either side of the center of the chain.}
\label{fig:TwoSpinCorrelators}
\end{figure}

Figure~\ref{fig:ising-entanglement}(a) shows a contour plot of dB squeezing in the
$(\alpha,t)$-plane, starting from a state that is fully polarized along the $x$ spin direction ($\varphi=\pi/2$, $\phi=0$). Increasing $\alpha$ leads to a slower and weaker creation of entanglement, but at the same time to a longer lifetime of the squeezed state. While there is a strong dependence of the spin squeezing on $\alpha$ for $\alpha<d$, where $d$ is the dimension of the system, the dependence
becomes weaker for $\alpha>d$.

Figure~\ref{fig:ising-entanglement}(d) shows dB spin squeezing in the
$(\varphi,t)$-plane
for $\alpha=3/4$. Maximal spin squeezing occurs for a tipping angle $\varphi=\pi/2$, implying that a
fully $x$-polarized initial state (or any fully polarized state in the $x$-$y$ plane) is the ideal choice for creating squeezed states under time
evolution. As the tipping angle is decreased to around $\varphi=\pi/4$,
spin squeezing persists and develops on a similar timescale to the $\varphi=\pi/2$ case, but reaches a smaller maximum value and disappears more quickly.
This can be understood 
by noting that correlations $\mathcal{C}^{az}_{ij}$ containing a $z$-component build up quickly
and reduce spin squeezing, as illustrated in Fig.\ \ref{fig:TwoSpinCorrelators}.
As $\varphi\rightarrow 0$, the squeezing approaches zero, and the time at which the small amount of squeezing is created  increases.

\textit{Entanglement entropy.}---The entanglement entropy between a subregion $\mc R$
and the rest of the system is given by
\be
S_{\mc R} = - \text{Tr}\left(\rho_{\mc R} \ln \rho_{\mc R} \right),
\ee
where $\rho_{\mc R}$ is the reduced density matrix of the subregion $\mc R$, obtained by tracing out all those
parts of the Hilbert space not associated with $\mc R$.
The entanglement entropy is
a measure of the total entanglement of bipartite pure states (although even un-entangled mixed states can have finite entanglement entropy) and is widely used in quantum information theory.

We will consider subregions ${\mc R}_{ij} \equiv\{i,j\}$, consisting of pairs of (not necessarily neighboring) spins $i$ and $j$. The reduced density matrix is given by
\begin{equation}
\rho_{{\mc R}_{ij}} = 4\sum_{a,b\in \{0,x,y,z\}}\expec{S^a_iS^b_j} S^a_i S^b_j,
\end{equation}
with the convention that $2 S^0_i$ is the $2\times 2$ identity matrix at lattice site $i$. From this expression it becomes obvious that the results for the one- and two-spin correlation functions in Sec.\ \ref{sec:ising-corrn} allow us to also obtain exact results for the entanglement entropy of subregions ${\mc R}_{ij}$.

Figures~\ref{fig:ising-entanglement}~(b,e) show $S_{{\mc R}_{ij}}$, where we have chosen ${\mc R}_{ij}$ as consisting
of the sites adjacent (one to the left and one to the right) of the center spin of a chain with 41 sites. Entanglement monotonically decreases for tipping angles away from $\varphi=\pi/2$.
This may be understood by considering that if we increase the tipping angle away from the fully $x$-polarized state, we increase the
$z$-component of the spin. Since components pointing along the $z$-axis are conserved under
the Ising time evolution, they will not contribute to the formation of entanglement. Hence,
as we move the tipping angle away from the fully $x$-polarized state at $\varphi=\pi/2$, we expect the
entanglement entropy to saturate at a lower value.

\begin{figure*}
\setlength{\unitlength}{1.0in}
\includegraphics[width=7in,angle=0]{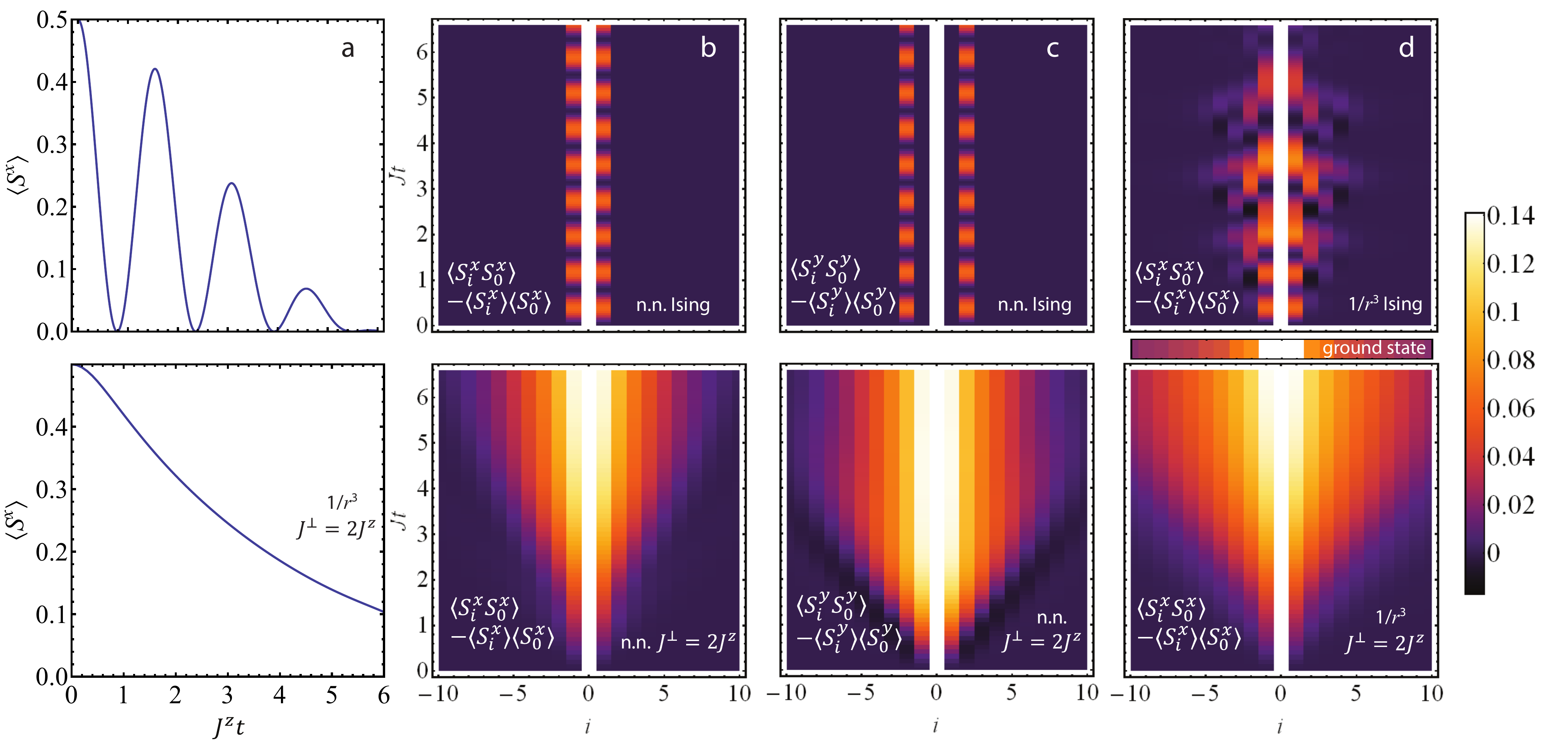}
\caption{
The dynamics of $\expec{S^x(t)}$  and correlation functions  for the Ising and XXZ models in one dimension. Panel (a)  shows  
$\expec{S^x(t)}$ versus $t$ for $\varphi=\pi/2$ and $\alpha=3$, for the  Ising model (top) and XXZ model with $J_\perp=2J_z$ (bottom). For $\varphi=\pi/2$,  $\expec{S^y}=\expec{S^z}=0$  due to symmetry considerations. The XXZ case is overdamped for this $J_\perp=2J_z$  case (and in general for large $J_\perp>J_z$) while the Ising dynamics damps only  when many different values of  $J_{ij}$ contribute to the dynamics. Other initial spin angles $\varphi$ are similar, but while $\expec{S^z}\ne 0$ remains constant  there is an additional precession or diffusion of the Bloch vector in the $x$-$y$ plane.
(b) Density plot of the two-point correlation function $\expec{S^x_i S^x_j}-\expec{S^x_i} \expec{S^x_j}$ versus $i-j$ (averaging over possible $j$'s to reduce finite size effects) and time for the nearest neighbor Ising model (top) and XXZ model with $J_\perp=2J_z$ (bottom).
(c) Same as panel (b), but showing $\expec{S^y_i S^y_j}-\expec{S^y_i}\expec{ S^y_j}$. (d) Same as panel (c), but for $\alpha=3$ interactions rather than nearest neighbor. The ground-state correlations are shown above the dynamics for the $J^\perp=2J^z$ case. 
 \label{fig:1d-correlations}}
\end{figure*}

Fig.\ \ref{fig:ising-entanglement}(b) shows a contour plot of the entanglement
entropy in the
$(\alpha,t)$-plane. The entanglement entropy evolution shows differences depending on whether
$\alpha$ is less or greater than $d/2$. For $\alpha$ substantially less than $d/2$ two well-separated timescales are apparent. On the shorter
 timescale (e.g. around $J^zt\approx 0.5$ for $\alpha=0.1$) a quasi-stationary state~\cite{vdW_NJP2013} of
intermediate entanglement strength is formed, while larger entanglement is built up  on the  longer
timescale (around $J^z t\approx 3$). The separation of scales is enhanced when increasing the number of lattice sites, or by  further reducing
$\alpha$. For  $\alpha>d/2$ no such separation of timescales is visible, i.e. the aforementioned plateau is not apparent. When $\alpha > d$ the
single- and two-spin correlation functions begin to show oscillatory behaviour [see Fig.\
\ref{fig:1d-correlations}(a)].  These oscillations become more pronounced for greater values of $\alpha$ and
can also be observed in the $(\alpha,t)$-plane for $\alpha \gtrsim  1.5$.
Linear and Renyi entanglement entropies can also be computed from Sec. \ref{sec:ising-corrn}'s correlation functions.
The
results obtained are qualitatively similar to those for the von Neumann entanglement entropy and are not shown here.

\textit{Concurrence.}---Concurrence was introduced in Ref.~\cite{Wootters_PRL1998} as an entanglement measure for
two-qubit systems that is valid for mixed states. It is defined as
\begin{equation}
 \mathcal{C}(\rho)\equiv \max\left\{ 0, \lambda_1 - \lambda_2 -\lambda_3 -\lambda_4 \right\},
\end{equation}
where $\lambda_1,\dots, \lambda_4$ are the square roots of eigenvalues (in decreasing order) of the non-Hermitian matrix $\rho_{\mathcal{R}_{ij}} \sigma^y_i \sigma^y_j \rho_{\mathcal{R}_{ij}}^* \sigma^y_i \sigma^y_j$. Concurrence is an entanglement measure valid for mixed states (whereas, for example,  entanglement entropy is an entanglement measure only for pure states), and its dynamics has recently been measured in trapped ion experiments~\cite{Jurcevic_arXiv_2014}.

While ${\mc S}_{{\mc R}_{ij}}$ measures the amount of entanglement between ${\mc R}_{ij}$ and the rest of the system (in the present case of a pure quantum state)~\footnote{The corresponding resource is the correlations that cannot be generated through quantum operations within each region and classical communication between them.}, the concurrence measures the amount of entanglement \textit{between} sites $i$ and $j$. 
Figure~\ref{fig:ising-entanglement}(c) shows a contour plot of the concurrence in
the ($\alpha,t$)-plane for a fully $x$-polarized, $\varphi=\pi/2$ initial state. The
concurrence reaches its maximum value for $d/2<\alpha<d$ and also persists for the longest time in this region. Similar to the
entanglement entropy in Fig.~\ref{fig:ising-entanglement}(b), the effect of oscillations that emerge in the
single- and two-spin correlation functions for $\alpha>d$ (here $d=1$) can be seen for $\alpha \gtrsim 1.8$. Fig.\
\ref{fig:ising-entanglement}(f) shows
the concurrence in the ($\varphi,t$)-plane for $\alpha=3/4$. Similar to other entanglement measures, the concurrence reaches its maximum
value  at $\varphi=\pi/2$. Away from this value, the growth of the concurrence is weaker and slower.

\textit{Summary of Sec.~\ref{sec:ising-entanglement}.}---For each of the entanglement measures studied here, we find a qualitative crossover in behavior as the exponent $\alpha$ is varied. The concurrence reaches a local maximum in 
$\alpha\in(d/2,d)$, while the spin squeezing goes from evolving at a rate essentially independent of $\alpha$ for $\alpha>d$ to a rate increasing with $\alpha$ for $\alpha<d$.  The
choice of the entanglement measure also determines the timescales at which the maximum entanglement is
reached. 
For all of the entanglement measures, an initial state in the $x$-$y$ plane ($\varphi=\pi/2$) is optimal for the creation of maximum entanglement under time evolution.

\section{1D correlations and entanglement\label{sec:1D-corrn-entanglement}}

In this section we investigate  
more general couplings with $J^\perp_{ij}\ne 0$, and we consider time scales much longer than the ones in which our perturbative short-time expressions of Sec.~\ref{sec:short-times} are valid.
To accomplish this we focus on one-dimensional systems, which are amenable to the adaptive t-DMRG~\cite{white:density_1992,vidal:efficient_2004,daley:time-dependent_2004,white:real-time_2004}; we use a Krylov-space implementation to treat the long-range interactions~\cite{2005AIPC..789..269M}. 
In addition to allowing us to investigate the $J_\perp\ne 0$ dynamics at relatively long times, 
the t-DMRG also allows us to compute entanglement entropies associated with arbitrary bipartitions of the system.  We will show that some features of the Ising limit, such as the types of correlations and the timescales on which they develop, 
are preserved in this model, while other features are modified, such as the range and propagation of the correlations.  These results give insight into when one can extrapolate the behavior of the Ising solution to more general cases, and help to elucidate the structure responsible for the solvability of the Ising limit.

Figure~\ref{fig:1d-correlations} compares and contrasts the dynamics of $\expec{S^x(t)}$ and some representative two-point correlation functions in the Ising limit and in the more general XXZ model. Figure~\ref{fig:1d-correlations}(a) shows the time-evolution of $\expec{S^x(t)}$ for $J_\perp=0$ (Ising, top) and $J_\perp=2J_z$ XXZ (bottom) models  with $1/r^3$ interactions, for an initial tipping angle of $\varphi=\pi/2$. (Similar results for $\expec{S^x(t)}$ were presented in Ref.~\cite{Hazzard}.)
The key difference between $\expec{S^x(t)}$'s evolution for the  Ising and  XXZ  cases is that the former oscillates substantially, while the latter is overdamped (although we note that for $\alpha \lsim d/2$ the Ising dynamics can become overdamped). The frequency spectrum of the dynamics provides a simple way to understand this behavior. In the Ising case, the frequency spectrum consists of all possible sums and differences of the coupling constants $J_{ij}$, as can be seen from Eq.~\eqref{eq:ising-Splus-expectation}. For $1/r^{3}$ (i.e. $\alpha=3$) interactions in 1D only a few frequencies are relevant over the timescale of interest, resulting in the pronounced oscillations. The same remains true for any interaction
that decays sufficiently fast in real space. In contrast, for the XXZ model, numerous other frequencies enter the spectrum (a continuous band of them emerges in the large-system limit), leading to an overdamped decay even for short-range interactions.

\begin{figure}[b]
\includegraphics[width=0.48\textwidth,angle=0]{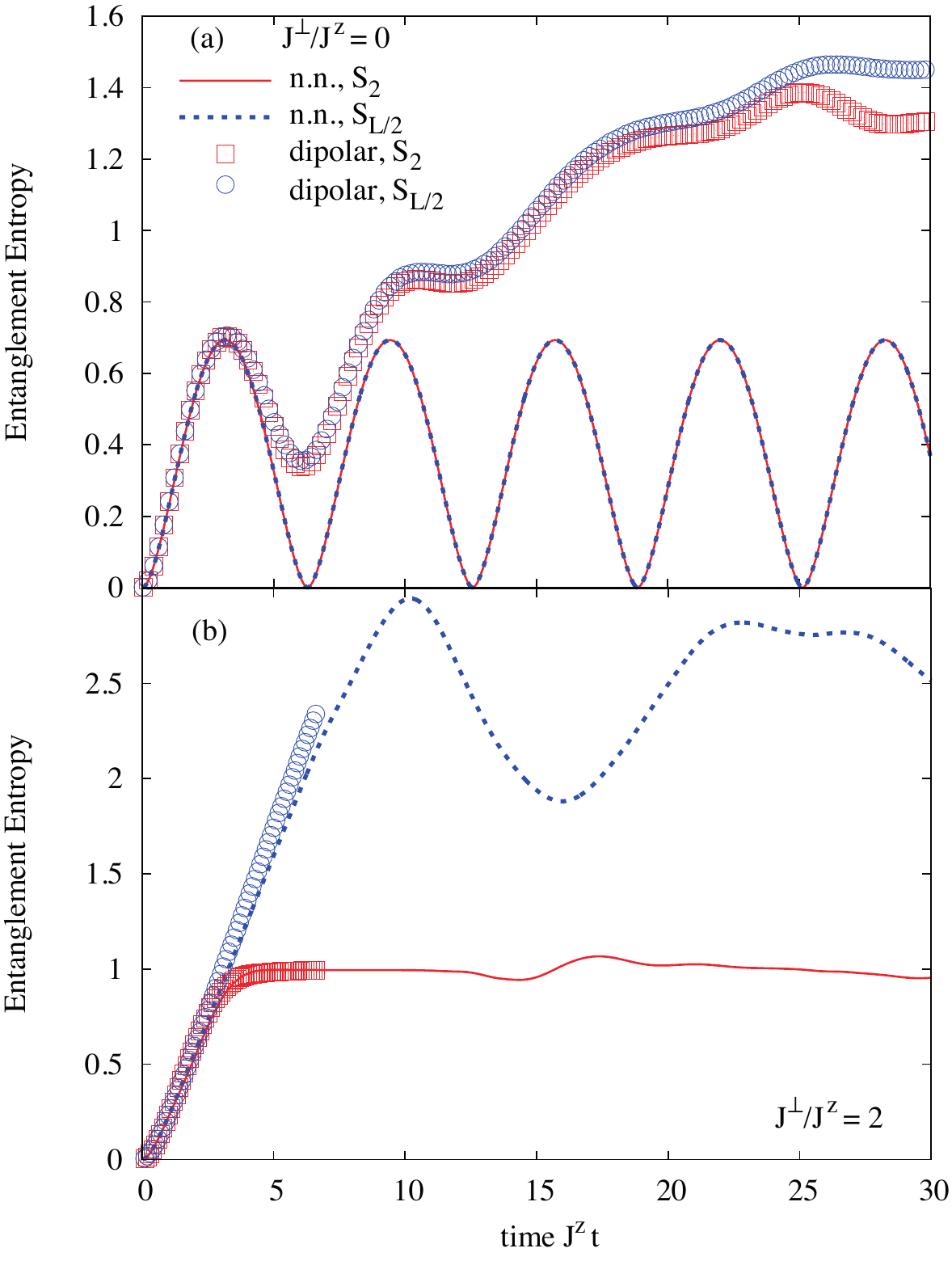}
\caption{Adaptive t-DMRG results for the time evolution of the entanglement entropy for XXZ chains with n.n. interactions (lines) and dipolar interactions (symbols). (a) $J^\perp/J^z=0$ (Ising); (b) $J^\perp/J^z = 2$ (XXZ). The plots display the entropy $S_2$ for a bipartition of size $\ell = 2$ at the edge of the systems (red) and $S_{L/2}$ at the center of the systems (blue). Note that in the Ising case the size of the bipartition does not play a role (nearest neighbor case) or is of minor relevance (dipolar case) so that the results are identical, up to boundary effects coming into play in the dipolar case. 
}
\label{fig:dynamics_entropy}
\end{figure}

Figure~\ref{fig:1d-correlations}(b) compares the Ising and XXZ dynamics of the spin correlation function $\mc G_{ij}^{xx}(t)$ for nearest neighbor interactions as a function of the site separation $i-j$ and time (horizontal and vertical axes, respectively). We show the results for nearest-neighbor interactions because they highlight a crucial distinction between Ising and general XXZ dynamics. Whereas these correlations in the Ising case propagate only to a distance determined by the range of the interaction, for the XXZ case, they can propagate further. This propagation in the XXZ case is evident in the  ``light cone''~\cite{calabrese:time_2006,Lauchli:2008dg,2009PhRvB..79o5104M,Cheneau:2012bh}  
structure~\footnote{The strict light cone~\cite{LiebRobinson_1972} must be generalized in the presence of interactions that are not finite ranged~\cite{HastingsKoma_2006,Eisert2013,
Richerme_arXiv_2014,Jurcevic_arXiv_2014}.} in the bottom panels of Figure~\ref{fig:1d-correlations}(b). The restriction of the Ising limit correlations to a finite range can be seen from the results in Eq.~\eqref{eq:ising-connected-corrn}, and has a simple physical explanation: In the Ising model, these correlations only build up due to direct interaction between two sites, and not due to propagation of excitations through repeated interactions. 
This distinction between Ising and XXZ dynamics is expected to hold in higher dimensions as well, where the Ising correlations still fail to propagate while the XXZ correlation will continue to do so, although the XXZ propagation likely will not remain ballistic.
Figure~\ref{fig:1d-correlations}(c) shows the same phenomenon for $\mc G^{yy}_{ij}$; the only distinction is that the maximum distance between correlated sites in the Ising limit is twice as large as the range of the interaction. This is because for a given value of $S^z_i$, interactions with a range $R$ drive the dynamics of sites ${i+R}$ and ${i-R}$, correlating them (through their mutual conditioning on the value of $S^z_i$) even though they are a distance $2R$ apart. In general, arbitrary correlators $\mc G^{ab}$ with $a,b\in \{x,y\}$ will be bounded to lie in this region of width $2R$. We note that the extra structure for the ${\mc G}^{xx}$ and ${\mc G}^{yy}$ correlators in the top panels of Fig.~\ref{fig:1d-correlations}(b,c) -- namely, that ${\mc G}^{xx}$ correlations extend to only half that distance and that ${\mc G}^{yy}$ correlations occur only at a separation of 2 and not at shorter ranges -- is a peculiarity of these specific correlation functions  for nearest-neighbor interactions. The ground-state correlations shown above the bottom panel of Fig.~\ref{fig:1d-correlations}(d) are included to illustrate that the correlations built up during the dynamics are, after very short times, larger than those in the ground state. 

 \begin{figure}[t]
\includegraphics[width=0.48\textwidth,angle=0]{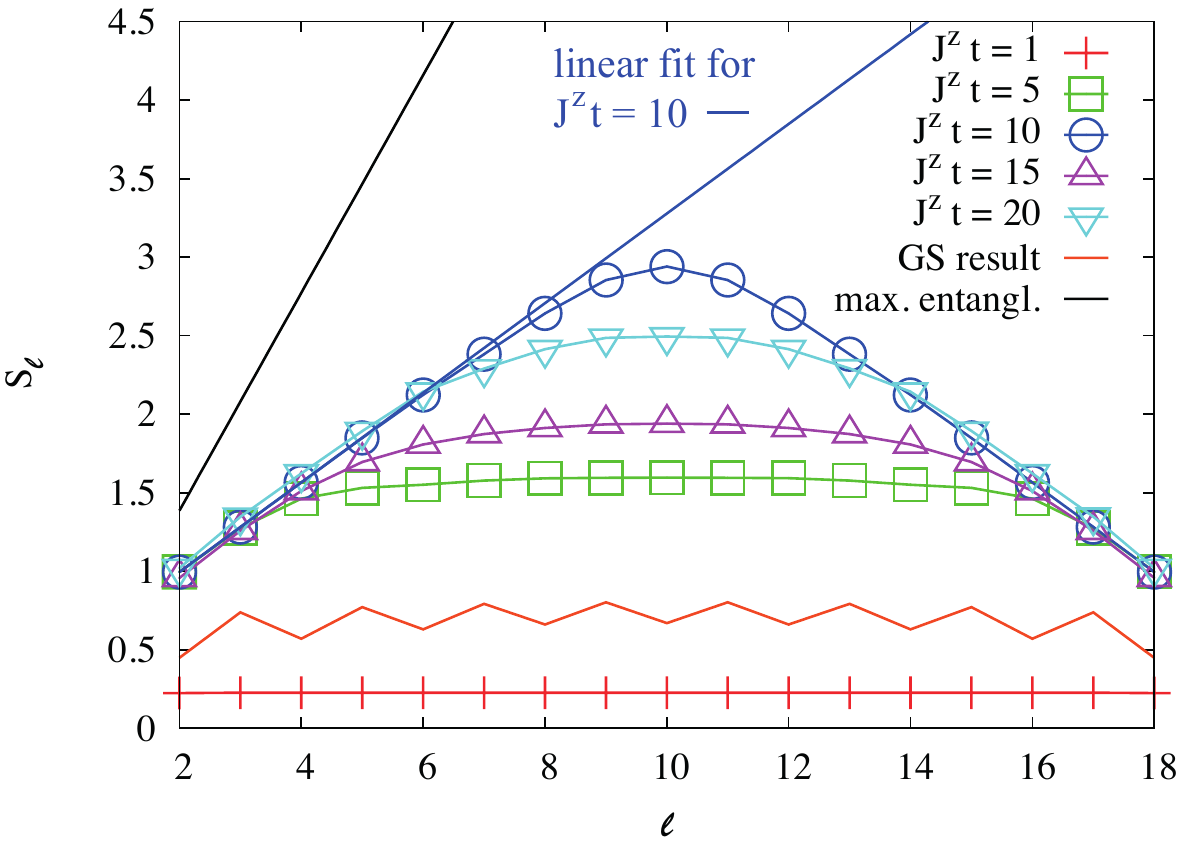}
\caption{ Adaptive t-DMRG results for the entanglement entropy of an XXZ chain with $J^\perp/J^z = 2$ and nearest-neighbor interactions as function of bipartition size at different instants of time. We compare to the ground state result for the same system with open boundary conditions in the red curve (``GS result").
Note the linear increase at the flanks (a fit is shown by the solid  blue line) which appears to reach a maximum slope in the course of the evolution. This is compared to the slope obtained for a maximally entangled state (labeled by ``max. entangl.").} 
\label{fig:entropy_bipartitions}
\end{figure}

Turning from correlations to entanglement, we quantify two types of entanglement generated during the dynamics: spin squeezing $\xi$, and the von Neumann entanglement entropy $S_{\ell}$ defined for a bipartitioning between sites $\ell$ and $\ell+1$.  Spin squeezing characterizes the temporal growth of metrologically useful entanglement, but since it is a spatially averaged quantity it yields no information about the spatial structure of the entanglement.  On the other hand, the entanglement entropy $S_{\ell}$ reveals both the temporal and spatial dependence of the entanglement, and is closely related to the numerical difficulty of simulating the dynamics with tensor-network based algorithms.

The spin squeezing dynamics for this model was calculated in Ref.~\cite{Hazzard}, where it was found to grow and reach a maximum on a timescale on the order of $1/J^\perp$ or $1/J^z$ and then to shrink and disappear; Fig.~\ref{fig:dynamics_entropy} shows, in contrast, typical behavior for the time-dependence of entanglement entropy. 
 At the beginning of the time evolution, the entropy $S_\ell$ grows independently of $\ell$.
 After a time that depends on the distance to the edge $\ell$, the entropy stops growing. This happens first at the edges, then the time for this saturation increases with distance to the edge. This gives rise  to an entropy that increases with $\ell$, shown in Fig.~\ref{fig:entropy_bipartitions}. 
This should be contrasted to the ground states of both gapped and gapless systems: in gapped systems there will be no dependence on $\ell$  for distances larger than a correlation length, and in  gapless systems there will be a logarithmic increase with distance to the edge~\cite{calabrese:entanglement_20014,Vidal:2003cn}.
As can be seen in Fig.~\ref{fig:entropy_bipartitions}, the entanglement in the nonequilibrium setup can be much larger than in equilibrium, even showing a linear dependence on the distance to the edge.  
At very long times, the entropy in the center of the chain  can oscillate in time, presumably due to finite-size/boundary effects, as seen in  Fig.~\ref{fig:dynamics_entropy}. 

The entanglement  entropy in the Ising limit is quite different. Rather than increasing linearly with distance $\ell$, the entanglement entropy is essentially independent of $\ell$ except near the edge. This is due to the lack of propagation of entanglement beyond a finite distance in short-ranged Ising models. The $1/r^3$ interaction in 1D is sufficiently short-ranged to be understood in this manner.

\begin{figure}[t] 
\includegraphics[width=0.48\textwidth,angle=0]{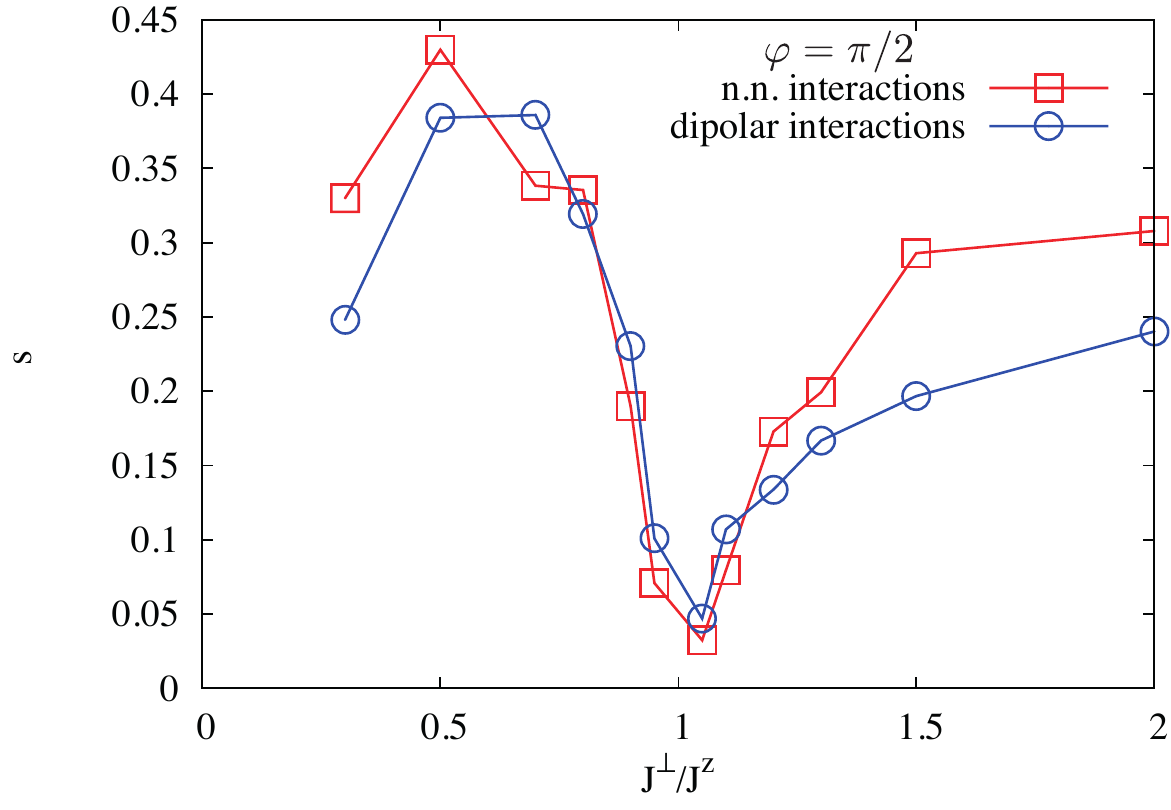}
\caption{ Maximal slope $s$ of the entanglement entropy's dependence on the distance to the system edge, defined by $S_\ell = s\ell$  near the edges of the system, obtained during the time evolution as a function of $J^\perp/J^z$ for $\varphi = \pi/2$. Red lines (squares): nearest neighbor interactions; blue lines (circles): dipolar interactions. Note that the slope (and hence the entanglement at the edges of the systems) appears to be larger for the nearest neighbor interaction than for the dipolar interactions when  $J^\perp > J^z$.}
\label{fig:slopes_entropy}
\end{figure}

The strong quantum fluctuations away from the Ising and the Heisenberg limits can induce strong entanglement, and it is natural to ask how close we can come to the maximum possible value.
We do this by considering the entanglement growth as a function of bipartition size at the edge of the system for sufficiently long times. 
As shown in Fig.~\ref{fig:entropy_bipartitions}, typically the entropy at the edge grows linearly, i.e., 
\be
S_\ell = s \ell,
\ee
 where the slope $s$ is the entanglement entropy density near the system edge. 
For spin-$1/2$ systems, a maximally entangled state is characterized by $S_\ell= \log (2^\ell)=\ell \log 2$ and thus $s = \log 2$.
Hence, a maximally entangled state is characterized by a volume law rather than an area law~\cite{calabrese:entanglement_20014,Vidal:2003cn}.  
Although our finding of a linearly increasing entropy indeed indicates a volume law at sufficiently long times, the value of $s$ obtained is always substantially lower than the maximally entangled $\log 2$.  
Figure~\ref{fig:slopes_entropy} shows $s$ as a function of $J^\perp/J^z$ for the quenches considered and for the cases where we could identify this volume law behavior~\footnote{For $J_\perp/J_z$ too small, the linear behavior is contaminated with finite size effects for the relevant times and system sizes we consider.}. 
As can be seen, a maximal value of $s\sim (2/3) \log 2$ is obtained for $J^\perp/J^z \sim 0.5$; it also increases to $\sim 1/3$ as  $J^\perp/J^z \to \infty$, and it approaches zero at the Heisenberg point $J^\perp \approx J^z$.  
The behavior is similar for both nearest-neighbor and dipolar interactions. 
Note that, surprisingly, for $J_\perp > J_z$, $s$ is actually larger for  nearest neighbor interactions than for long-range interactions.

\section{Universality out of equilibrium\label{sec:univ-out-of-eq}}

In this section we demonstrate how the considered dynamic protocol can reveal non-equilibrium universality -- dynamics having characteristic features that are insensitive to changes of the microscopic model.  The existence of universal behavior in non-equilibrium systems is especially interesting for two reasons. First, it is non-perturbative; any perturbation theory in the Hamiltonian or time will depend on the microscopic details of the perturbation. Second, while universality in equilibrium systems is explained by one of the most fundamental tools in physics, the renormalization group (RG), the required concepts and tools for generalizing it to far-from-equilibrium systems have yet to be developed (though some progress has been made in this direction, see for example Refs.~\cite{degrandi:quench_2010,degrandi:quench-sine-gordon_2010,gritsev:universal_2010,mathey:light_2010,
mathey:dynamics_2011,mitra:mode-coupling_2011,mitra:thermalization_2012,mitra:time_2012,vosk:many-body_2012,karrasch:luttinger-liquid_2012,sarkar:perturbative_2013,dallatorre:universal_2013}).

We consider the dynamics of an initial state with  tipping angle $\varphi=\pi/2$, evolving under a one-dimensional XXZ Hamiltonian that deviates only slightly from the SU(2) symmetric point, i.e. $| J^\perp/J^z-1|\ll 1$. However, although the explicit formulas will change, the existence of the universal singularity persists for arbitrary $\varphi$. Since the initial state is a ground state when $J^{\perp}=J^z<0$, the dynamics can be understood as a small quench starting from the ground state at the SU(2) symmetric point, as in Fig.~\ref{fig:dynamics-cartoon}(b).  We  consider the case $J^{\perp}/J^z>1$ so that the ground state of the XXZ Hamiltonian governing the time-evolution has gapless excitations. Due to the model's time-reversal invariance the overall sign of the Hamiltonian is irrelevant for the dynamics, although it is necessary that $J^\perp$ and $J^z$ have the same sign.
Throughout this section  we assume a ferromagnetic coupling $J^{\perp}<0$ where the arguments are clearest.   For simplicity we restrict our attention in the exact numerics to  nearest neighbor interactions, $J^{\{\perp,z\}}_{ij} =J^{\{\perp,z\}} \lp \delta_{j,i+1}+\delta_{j,i-1}\rp$.

Given that we are in  the ``small'' quench regime,  it is plausible that a low-energy description of the final Hamiltonian is appropriate to account for the dynamics.
We first determine the low-energy theory near the SU(2) symmetric point, $J^\perp=J^z=-J$. We do this by writing the action as $S=S_0+\delta S$ with $S_0$ the SU(2) symmetric point action and $\delta S$ the deviation from it.  At the SU(2) symmetric point the low-energy excitations are spin waves (delocalized single spin flips) that have a quadratic dispersion. By symmetry, the corresponding action is
\be
S_0 &=& J \int \! d\omega \, dk \,\lp a \omega^2 + b k^4 \rp |\phi_{k,\omega}|^2,
\ee
where $\phi$ is the bosonized field~\footnote{Specifically, $\phi$ is defined so that its spatial derivative is the fermion density obtained by a Jordan-Wigner transform.}, $a$ and $b$ are model-dependent constants whose values are unimportant here, and higher-order RG-irrelevant terms have been dropped. 
This form is expected since at the SU(2) Heisenberg point, the elementary excitations are non-interacting spin waves, behaving as free bosons with a $k^2$ dispersion. 
To utilize this low-energy theory, we need to relate physical observables to the field $\phi$. For initial states in the $x$-$y$ plane ($\varphi=\pi/2$), the field $\phi$  measures the in-plane direction of the (coarse-grained) local magnetization 
\be
S^x
&=& N/2 \cos(\phi). \label{eq:phi-from-bosonization}
\ee

The perturbation from this point when $0<\delta J\equiv J^z-J^\perp \ll 1 $ is
\be
\delta S &=& \gamma \delta J \int \! d\omega \, dk \,k^2 |\phi_{k,\omega}|^2
\ee
where $\gamma$ is an unimportant model-dependent proportionality constant, and we have again  neglected  RG-irrelevant terms. In the microscopic picture, the $\delta J$ coupling corresponds to repulsive interactions between the spin waves. Finally, the total action $S$ to leading order in $\delta J$, $k$, and $\omega$ is 
\be
S &=& \frac{K}{\pi} \int \! d\omega \, dk \, \lp \frac{1}{v} \omega^2 + v k^2 \rp |\phi_{k,\omega}|^2,\label{eq:LL-from-perturbation}
\ee
the Luttinger liquid Hamiltonian with Luttinger parameter $K= \sqrt{a\gamma \delta J/J}$ and velocity $v=\sqrt{\gamma \delta J/(aJ)}$.  Both $K$ and $v$ have a singular $\sqrt{\delta J}$ dependence. (This equation can also be obtained by a Bogoliubov expansion for small $\delta J$.)  We will return to the universality of these predictions later.

Now we compute the dynamics for the initial state evolving under Eq.~\eqref{eq:LL-from-perturbation}, assuming this low-energy theory suffices for this computation. Using the relation Eq.~\eqref{eq:phi-from-bosonization} and the fact that Eq.~\eqref{eq:LL-from-perturbation} is Gaussian in $\phi$, we have $\expec{S^x(t)}=N/2 \exp(-1/2\expec{\phi^2})$. Following Ref.~\cite{Bistritzer:intrinsic_2007}, we find three qualitatively different types of behavior depending on the time,
\begin{equation}
\expec{S^x(t)} \propto
\begin{cases}
\lb 1-(\delta J t)^2 \rb \hspace{0.1in} & \text{for $t\ll 1/\delta J$},\\
  e^{-t/\tau} \hspace{0.45in}& \text{for $1/\delta J \ll t \ll L/v$},\\
 e^{-(t/\tau_{\rm fs})^2} \hspace{0.15in} & \text{for $t \gg L/v$}.
\end{cases}
\label{eq:universal-dynamics-3-regions}
\end{equation}
Here $\tau= 4 a K^2/(\pi^2 v  )$, $a$ is the lattice spacing, $\tau_{\rm fs}^2= \tau L/(2v)$, and $L$ 
 is the chain length (assumed to be large). The crucial observation is that near the SU(2) point these parameters follow a universal power law. In particular
\be
\tau &=& A \sqrt{J/\delta J^3}\label{eq:tau-univ}
\ee
(with some nonuniversal numerical prefactor $A$). This leads to a universal spin coherence decay time  in the intermediate-time regime $1/\delta J \ll t \ll L/v \sim L/\sqrt{\delta J J}$ (thus the appropriate separation of scales for this regime to exist is $L\gg \sqrt{J/\delta J}$). Physically, the spin dynamics in this regime results from dephasing of the initially populated Luttinger liquid excitations~\footnote{For $\varphi\neq 0,\pi$ the evolution of $S^x$ corresponds to a decay from dephasing of the individually oscillating excitations. The frequency of the oscillations is approximately given by $\delta J \expec{S^z}=\delta J \cos \varphi$. The resulting damped dynamics of $\expec{S^x}$ is described by the Luttinger liquid Hamiltonian Eq.~\eqref{eq:LL-from-perturbation}, with renormalized bosonic density $\rho_0 = (1-|\cos\varphi|)/(2 a)$. Note that the states at $\varphi=0,\pi$ are exact eigenstates of the XXZ model and consequently do not evolve dynamically.}.

The intermediate-time result demonstrates the non-perturbative nature of the dynamics, since $e^{-t/\tau}$ with $\tau \propto \sqrt{J/\delta J^3}$ is non-analytic in $J^z t$ and $J^\perp t$. Consequently no order of perturbation theory in the bare Hamiltonian Eq.~\eqref{eq:XXZ-model} would reproduce these results. We note that  if $\expec{S^x(t)}$ is smooth, as it appears to be, this non-analyticity is  signaled as a divergence somewhere in the complex $J^z t$  or $J^\perp t$ plane. In this regard, our results resemble the dynamic singularities found on the real-time axis in Ref.~\cite{heyl:dynamical_2013}.

The derivation of our intermediate-time dynamics also suggests that the results are universal, in two senses. The first sense of universality is that the dynamics depends only on the Luttinger parameter $K$ and velocity $v$. Contrast this with the short-time behavior in Eqs.~\eqref{eq:short-time-corrns}, which depends on the detailed couplings of the model. Irrelevant perturbations to the model (such as a second-nearest-neighbor coupling) affect the short-time dynamics but not the intermediate-time dynamics, once written in terms of the Luttinger parameter. Physically, at these times, any high-energy degrees of freedom have had time to dephase and stop contributing to $\expec{S^x(t)}$ so that only the low-energy excitations governed by Luttinger liquid theory contribute. The second sense of universality is a further insensitivity to microscopic perturbations, and is related to the fact that we are quenching from the SU(2) point. Here, the Luttinger parameters and thus  $\tau\propto K^2/v$ is a universal 
power law proportional to  $\sqrt{J/\delta J^3}$. 

\begin{figure}
\setlength{\unitlength}{1.0in}
\includegraphics[width=3.45in,angle=0]{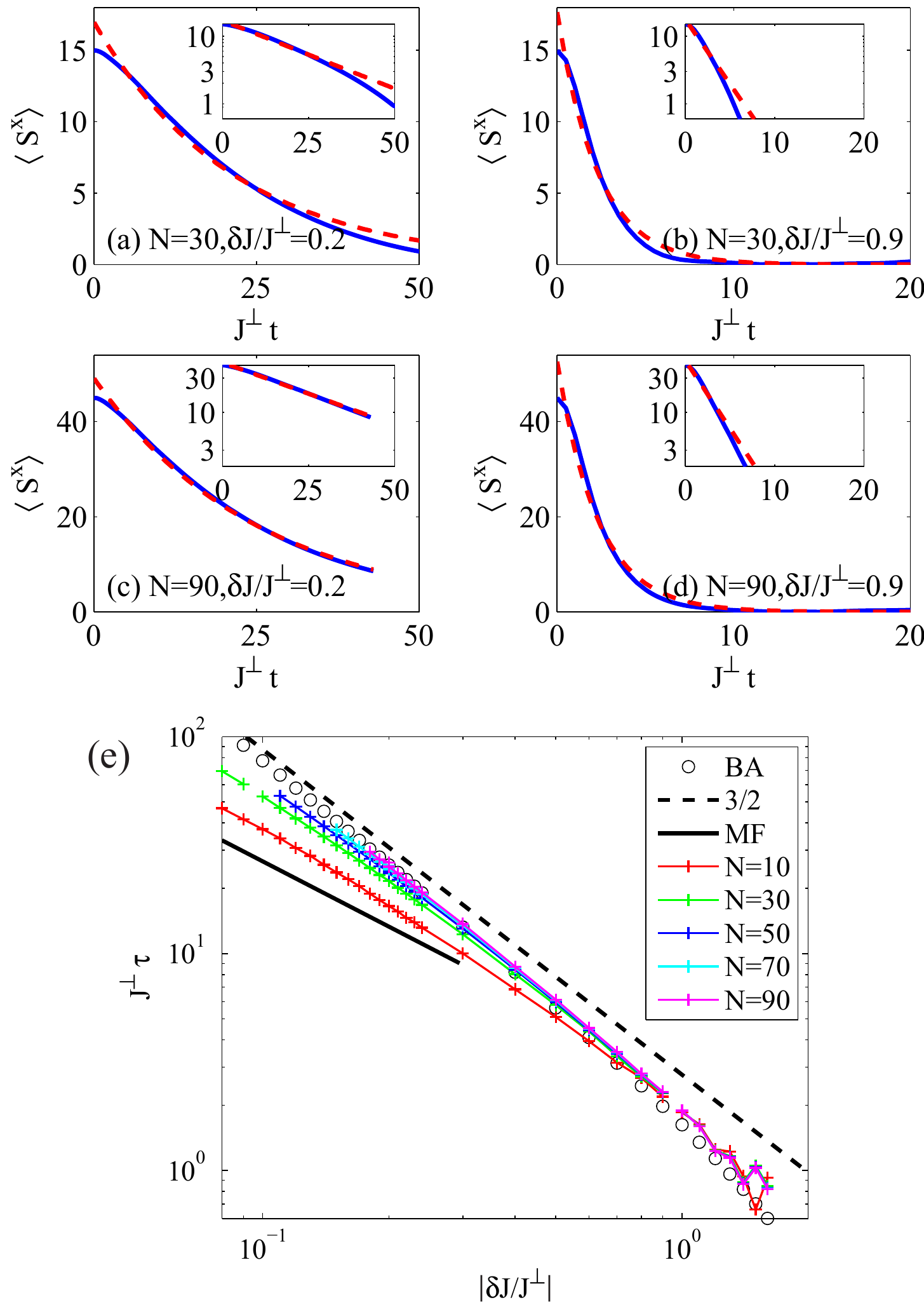}
\caption{ 
Universal scaling in far-from-equilibrium dynamics. (a-d) Dynamics of $\langle S^x(t)\rangle$ for two different values of the quench parameter $\delta J = J_\perp - J_z$ and of the system size $L$. Solid lines are t-DMRG results, and dashed lines are exponential fits $A_0 e^{-t/\tau}$ over the appropriate intermediate-time
region (see main text). (e) Decay time $\tau$ as a function of the quench parameter $\delta J$. Open circles: universal result, Eq.(34), with Luttinger liquid parameters determined by Bethe ansatz (BA). Dashed line: universal scaling $ J \tau \propto (J/\delta J)^{3/2}$, valid for $|\delta J/J_\perp| \ll 1$. Joined symbols: t-DMRG results for various system sizes from $L = 10$ to $L = 90$. Large systems converge to the universal Luttinger liquid prediction, while small systems follow a mean-field (MF) behavior $\tau \propto \delta J^{-1}$.
 \label{fig:univ-far-of-equilibrium}
 }
\end{figure}

Although the preceding analysis argued for universality based on  Luttinger liquid theory, it does not determine the range of $J_\perp-J_z$ and system sizes $L$ for which its  conclusions are valid.
To evaluate these parameter regimes, we use t-DMRG, implemented with time-evolving block decimation (TEBD) in the ALPS package~\cite{wall:tebd-2009,bauer:alps_2011}.
Figures~\ref{fig:univ-far-of-equilibrium}(a--d) show the dynamics of $\expec{S^x(t)}$ for $\varphi=\pi/2$ and several quenches of chains with lengths up to $L=90$ sites, differing in the final value of $\delta J$. Exponential decay is evident at intermediate times when this regime is accessible to the numerics, in accord with the Luttinger liquid prediction in Eq.~\eqref{eq:universal-dynamics-3-regions}. (As an aside we note that for the specific model studied here, it may be possible to employ recently developed exact analytic methods~\cite{liu:quench_2013} to study much of the time dynamics.)

Figure~\ref{fig:univ-far-of-equilibrium}(e) shows the dependence of the exponential decay exponent $\tau$ on $\delta J$. We determine $\tau$ by fitting~\footnote{Namely, we fit to times
\unexpanded{$0<t<t_{\text{fit}}$}
where
\unexpanded{$t_{\text{fit}}$}
defined by
\unexpanded{$\expec{S^x(t_{\text{fit}})}=\expec{S^x(t_{\text{fit}})}/4$}.
In addition to long times, short times should also in principle be excluded form the fit, but for all of the
$t_{\text{fit}}$
presented here, the contribution from short times is negligible.}  the intermediate-time behavior of $\expec{S^x(t)}$  to an exponential decay for each $\delta J$.  The results confirm our  prediction [Eq.~\eqref{eq:tau-univ}] that $\tau \propto \sqrt{J/\delta J^{3}}$ for small $\delta J$.

In order to reach this universal regime, one must go to times long compared to the microscopic cutoff time, which is $O(\delta J^{-1})$. However, in small systems, finite size effects begin to play a role before reaching this universal regime. Thus the characteristic scale is set by  $\delta J t$, coinciding with the mean-field timescale $\tau \sim \delta J^{-1}$, which indeed appears to be approached for small systems, as shown in Fig.~\ref{fig:univ-far-of-equilibrium} (solid line).

Going beyond the limit of universal power law scaling of $\tau$ (i.e., small $\delta J$), we can compute the expected $\tau=4K^2a/(\pi^2  v)$ by using Luttinger liquid parameters $K$ and $v$ that can be determined by the Bethe ansatz~\cite{takahashi:thermodynamics_1999}. We see that even for some $\delta J$ large enough that the $\delta J^{-3/2}$ scaling breaks down, $\tau$ is still related to the Luttinger liquid $K$ and $v$ in the appropriate regime, confirming universality in the first sense described above.  

We note that we have concentrated on $\{J_\perp,J_z\}<0$, but for $J_z>0$ (or equivalently $|\delta J/J^\perp|>1$) the universal behavior breaks down even in this first sense: the numerical calculations   show a significant deviation
from the Luttinger liquid theory predictions. In this $J^z>0$ regime,  the initial state is far away from the actual ground state of the system. (For $J^z>|J^\perp|$, the ground state is qualitatively different, becoming antiferromagnetic rather than ferromagnetic.) As a consequence the dynamics is probably not well described by the Luttinger liquid theory. 
Further analytic and numerical calculations are needed in order to elucidate the long time behavior in this regime and determine the possible emergence of a
new universal behavior. Some calculations exploring this were presented in Ref.~\cite{barmettler:quantum_2010}.
 
We conclude this section with an outlook based on our findings. 
Although we focused on one-dimensional systems, our results suggest that a universal behavior could be observed in higher-dimensional systems as well, where low-energy effective theories may continue to describe the dynamics after a small quench. However, even if this  holds in higher dimensions, a suitable framework is still lacking for two reasons: (1) it is harder to identify the low-energy theories and (2) the dynamics of these low-energy theories is significantly harder to calculate -- it is frequently intractable to present techniques. 
Ideas suitable for a more general quench that are based on real-time renormalization group have been proposed and developed in Refs.~\cite{mitra:mode-coupling_2011,dallatorre:dynamics_2012,mitra:thermalization_2012,
mitra:time_2012,mitra:correlation_2013}, and for some quench scenarios conformal field theory~\cite{calabrese:time_2006}, truncated Wigner approximation~\cite{mathey:dynamics_2011} and semiclassical approaches have been employed~\cite{dallatorre:universal_2013}.

An interesting aspect of our results is that they provide a route to observing Luttinger liquid physics in ultracold lattice spin systems that does not require temperatures below the reach of ongoing cold-atom experiments (as opposed to more easily accessible  Luttinger liquid physics associated with the density degree of freedom). In contrast to the current dynamic proposal, the observation of Luttinger liquid physics in spin systems in equilibrium requires cooling to (currently unfeasible) temperature scales much smaller than the magnetic interaction scale, which can be $\sim 1$~nK for superexchange-based implementations in ultracold atoms~\cite{bloch_many-body_2008}. This is true even for proposals that rely on nonequilibrium probes of equilibrium states, such as Ref.~\cite{pollmann:linear_2013}.
In our procedure, the initial state is easily prepared, and the dynamics is used to probe the Luttinger liquid physics. Observing the dynamics governed by the Luttinger liquid physics requires reaching timescales that are long compared to those determined by the magnetic interaction. However, even for a superexchange scale typical in cold-atom experiments ($\sim 1$~nK), the corresponding timescale is $\sim 50$~ms, and ought to be experimentally accessible.

\section{Physical realizations \label{app:phys-real}}

Specific instances of the XXZ Hamiltonian in Eq.~\eqref{eq:XXZ-model} and the dynamical protocol discussed in this paper can be implemented in a variety of atomic, molecular, optical, and solid-state systems, including but not limited to polar molecules \cite{yan:observation_2013,hazzard:many-body_2014}, trapped ions \cite{kim:quantum_2010,lanyon:universal_ion_2011,britton:engineered_2012,islam:emergence_2013,Richerme_arXiv_2014},  Rydberg atom ensembles \cite{Weimer_NaturePhys_2010,Schaub_Nature_2012,Low_JPhysB_2012}, 
magnetic quantum gases~\cite{depaz:nonequilibrium_2013}, neutral atoms in optical lattices \cite{trotzky:time-resolved_2008,Greif_Science_2013}, alkaline-earth-atom optical lattice clocks \cite{swallows:suppression_2011,martin:quantum_2013}, tilted Bose-Hubbard models in optical lattices \cite{Simon_Nature_2011}, magnetic defects in solids (e.g. nitrogen-vacancy centers in diamond) \cite{Prawer_Science_2008} and quantum magnetic materials in solid state physics. For these solid state systems, Refs.~\cite{auerbach:interacting_1994,Sachdev_Book_1999,
lacroix_introduction_2011,sachdev_quantum_2008} give theoretical overviews, and Ref.~\cite{kaur:decay_2013} is just one example of recent experiments probing quench dynamics similar to that considered here. The models are also important for molecular aggregates~\cite{alvarez:quantum_2013,alvarez:localization_2011,alvarez:nmr_2010} and energy transport in large organic and biomolecules~\cite{vragovic:frenkel_2003,combescot:microscopic_2008,
litinskaya:rotational_2012,xiang:tunable_2012,
hoyer:realistic_2013,saikin:photononics_bio_2013}. Experiments in many of these systems have measured the same or closely related quench dynamics to that considered herein.
The diversity and scope of these various implementations is enormous, but here we attempt to give a brief review of some of the most promising implementations, including their unique benefits and limitations. Table~\ref{tab:systems-overview} of Sec.~\ref{sec:models} summarizes the relevant properties (e.g. spin-coupling structure, energy scales, coherence times, etc.) of several of the systems described below.

\subsection{Quantum magnetism in polar molecules}

Polar molecules pinned in optical lattices have numerous desirable qualities for the  realization of spin
Hamiltonians. They have a variety of internal degrees of freedom --- hyperfine, rotational, vibrational, and
electronic --- spanning many orders of magnitude in energy, all of which in principle can be used to encode a
spin degree of freedom~\cite{barnett:quantum_2006,micheli:toolbox_2006,
wall:hyperfine_2010,gorshkov:quantum_2011,gorshkov:tunable_2011}. Moreover, since long-range dipole-dipole interactions do not require wave-function overlap, they persist in the deep-lattice limit (where tunneling is negligible) and are typically quite strong: $J/h\sim 10^2$ Hz at $\sim 500\rm{nm}$ lattice spacings for KRb, a relatively weakly interacting molecule; for LiCs the interactions can be a hundred times larger. When rotational levels encode the spin degree of freedom, an Ising term can be induced by permanently polarizing the molecules with modest electric fields, while spin-exchange terms ($\propto J_{\perp}$) of comparable strength can arise from resonant microwave photon exchange. The recent experiments reported in Refs.~\cite{yan:observation_2013,hazzard:many-body_2014} have demonstrated the existence of $J_z=0$ exchange dynamics  in
 KRb molecules first prepared in the ro-vibrational ground state,  pinned in a 3D optical lattice, and excited
using the  quench protocol described in this paper. Those experiments have additionally  demonstrated the capability to control the interaction strength by choosing different rotational levels to represent the spin degree of freedom. In
the future, the addition of DC electric fields and microwave or Raman dressing is expected to enable the simulation of the  more general XXZ spin models that we consider here, and more~\cite{gorshkov:tunable_2011,yao:topological_2012,yao:realizing_2013,gorshkov:kitaev_2013}. Current challenges in quantum simulation with polar molecules include the production of the molecules themselves (so far only KRb has been successfully produced in its rotational, vibrational, and hyperfine ground state at motional temperatures near quantum degeneracy), achieving sufficiently high phase space density (i.e. approaching unit filling in a lattice) \cite{Freericks_PRA_2010,chotia_long-lived_2012}, and understanding any relevant decoherence mechanisms.

\subsection{Quantum magnetism in trapped ions}

In existing implementations of trapped-ion quantum simulators, the ions are laser cooled to form either one-dimensional (rf-Paul trap)~\cite{kim:quantum_2010} or two-dimensional~(Penning trap) \cite{britton:engineered_2012} crystals. Many of the same properties that have made trapped ions a leading platform for quantum computation make them especially promising and versatile as simulators of models of quantum magnetism. First, they possess long-lived hyperfine states, which form the spin degree of freedom in all quantum simulations using trapped ions to date. While the spin degrees of freedom of two $\gtrsim\mu$m separated ions do not interact directly on experimentally relevant timescales, the spin of an individual ion can be coupled to its motion via a spin-dependent force. Experimentally, this is induced by off-resonant stimulated Raman transitions. By modulating this force at a frequency $\mu$, virtual phonons are excited which mediate relatively large ($J\lesssim 10$kHz) spin-spin couplings. These spin-spin couplings inherit the
nonlocal structure of the phonon modes, and hence are generically long-ranged; depending on how far $\mu$ is detuned from the various phonon modes of the ion crystal, the coupling between two ions falls off in space as an approximate power law of the distance between them, with exponent $0\leq\alpha\leq3$. To date, experiments have successfully implemented Ising ($J_{\perp}=0$) \cite{kim:quantum_2010,britton:engineered_2012,islam:emergence_2013} and XX ($J_{z}=0$) \cite{Richerme_arXiv_2014,Jurcevic_arXiv_2014} models both in and out of equilibrium, but in principal these systems can be used to realize generic anisotropic spin models ($J_x\neq J_{y}\neq J_z$) \cite{porras:effective_2004}. Current challenges in trapped ion quantum simulation are mainly related to scalability; pushing to larger system sizes adversely affects the ratio of the spin-spin interaction timescales to the system lifetime, and may eventually require working in cryogenic environments. We note that there are a variety of possible
complications due to the production of real phonon excitation during non-equilibrium dynamics in these systems \cite{Wang2012}; the importance of these effects is only partially understood.

\subsection{Quantum magnetism in Rydberg atoms}

Very similar  
to polar molecules, Rydberg atoms possess a strong dipole-dipole interaction. As the dipole moment is  limited only by the size of the Rydberg atoms, their mutual interaction can be many orders of magnitude stronger than all other interactions between neutral atoms or molecules. If a pair of two excitations in Rydberg atoms is resonant with dipole-allowed transitions to other two-excitation states, then the dipolar coupling between them results in an anisotropic long range $1/r^3$ interaction. If this so-called F{\"o}rster process is off-resonant,  this gives rise to  the well-known attractive or repulsive $1/r^6$ van der Waals interaction. The F{\"o}rster resonance can be tuned and experimentally controlled in fast pulse sequences by electric fields that act differently on the involved Rydberg states, thus enabling one to control the detuning between the two-excitation states.  Another experimental tool is microwave pulses, which can be used to induce oscillating dipole moments. This is an ideal tool in Ramsey-like pulse sequences. It is important to note that the lifetime of Rydberg atoms is  limited typically to $\sim$10 microseconds. However, coherent driving as well as interaction timescales  have been shown to reach even the GHz level~\cite{huber:GHz_2011,baluktsian:evidence_2013}  which is ultimately limited only by the Kepler frequency -- i.e. the energy splitting between adjacent Rydberg states. On these time scales it is not even necessarily required to work with a degenerate ultracold gas, as even at elevated temperatures the system behaves as a frozen gas on the microsecond time scale.

Strongly interacting frozen Rydberg gases have been shown to be able to quantum simulate the ground states of spin Hamiltonians. Following the theoretical prediction of a quantum phase transition from a paramagnetic to  crystalline
phase~\cite{weimer:quantum_2008} the universal scaling behavior in the quantum critical regime was experimentally investigated and compared to
\textit{ab initio}
and mean-field
calculations~\cite{universal:loew_2009}. Spatially
ordered ground states have been observed by various methods~\cite{schwarzkopf:imaging_2011,schwarzkopf:spatial_2013}, including in Ref.~\cite{Schausz:obervation_2012}, which employed an underlying optical lattice.
Ramsey sequences have been used to monitor the nonequilibrium dynamics and interaction~\cite{anderson:dephasing_2002,butscher:atom-molecule_2010,nipper:highly_2012}. Quenches to F{\"o}rster resonant interaction were investigated as early as 1998~\cite{anderson:resonant_1998,mourachko:many-body_1998}, and they have recently been used to investigate spatial diffusion processes~\cite{guenter:observing_2013}.

\subsection{Quantum magnetism of neutral atoms in optical lattices via dipolar interactions}

Strong magnetic interactions between atoms with unpaired electrons give rise to mechanical effects in a trapped quantum gas that were observed in Ref.~\cite{stuhler:observation_2005}. They also result in strong dipolar relaxation processes~\cite{hensler:dipolar_2003}, which have been used for demagnetization cooling~\cite{fattori:demagnetization_2006}. Ref.~\cite{hensler:dipolar_2003} noted that the dipolar interaction can lead to an exchange term, 
but in  free space this is always accompanied by a magnetization relaxation term.  It was then demonstrated that in a dipolar lattice gas the relaxation of the  magnetization can be suppressed~\cite{depaz:resonant_2013}, in which case the remaining terms of the dipolar interaction are of the form Eq.~\eqref{eq:XXZ-model} 
with $J^\perp/J^z$ fixed by the  nature of the dipolar interaction. Recently, an experiment in the many-body limit with singly and doubly occupied lattices sites was performed and showed evidence for coherent inter-site spin exchange dynamics~\cite{depaz:nonequilibrium_2013}. So far these experiments have been performed with Chromium atoms, which have a magnetic moment of 6 $\mu_B$, where $\mu_B$ is the Bohr magneton, and a ground state manifold with a spin $S=3$ degree of freedom. Upcoming experiments with Erbium (7 $\mu_B$) and Dysprosium (10 $\mu_B$) promise even stronger couplings~\cite{aikawa:bose-einstein_2012,lu:strongly_2011}.

\subsection{Quantum magnetism of neutral atoms in optical lattices via superexchange}

Neutral atoms in optical lattices furnish natural realizations of Bose or Fermi Hubbard models \cite{Jaksch_PRL_1998}. Much like the electrons in real materials, at unit filling and in the strongly interacting limit these atoms form Mott insulators with  one particle per lattice  site \cite{Greiner_Nature_2002,Jordens_Nature_2008,Schneider_Science_2008}. If each atom can be in one of two hyperfine (spin) states, virtual excitations into states away from unit filling, i.e. with holes and double occupancies, mediate spin-spin interactions between neighboring atoms that are ferromagnetic (antiferromagnetic) for bosons (fermions). By choosing internal states that have different scattering properties, or by using hyperfine-state-dependent lattices, or both, it is possible to tune the relative strength of $J_z$ and $J_{\perp}$\cite{duan:controlling_2003}. These superexchange interactions have been probed in bosonic Rb atoms via non-equilibrium dynamics in double-well arrays \cite{trotzky:time-resolved_2008}, and 
recently   in equilibrium in anisotropic  lattice geometries  using low-temperature  fermionic Mott insulators \cite{Greif_Science_2013}. However, due to the very small superexchange energy scales ($J\lesssim 100$Hz, often much less), studying equilibrium properties of fermionic systems in the temperature regime where long-range antiferromagnetic order exists remains elusive. In alkaline-earth atoms, the independence of the scattering properties on the nuclear spin $I$ allows for superexchange models with SU$(N=2I+1)$ [rather than SU$(2)$] symmetry \cite{Wu2003,Gorshkov_NaturePhys_2010,cazalilla2009,Rey_arXiv_2014}. Recently, direct and indirect signatures of SU($N$) symmetry in $s$- and $p$-wave collisions  have been reported in Refs.~\cite{Stellmer2011,Taie_NaturePhys_2012,Zhang_arXiv_2014,Scazza_arXiv_2014}. Signatures of a superexchange processes  respecting this enhanced SU$(N)$ symmetry have not been demonstrated experimentally, however the required low-entropy SU$(N)$ Mott-insulators with $N=6$ 
have been recently created \cite{Sugawa_NaturePhys_2011,Taie_NaturePhys_2012,Scazza_arXiv_2014}.

\subsection{Quantum magnetism of alkaline-earth-atoms in optical lattice clocks}

Alkaline-earth atoms trapped in 1D and 2D optical lattices can realize spin models in appropriate limits. Here the role of lattice sites is played by single-particle quantized motional eigenstates (e.g., harmonic oscillator levels along the ``tubes" or ``pancakes" of this lattice), and the spin degrees of freedom are encoded in two electronic states  (clock states) \cite{rey:many-body_2009,Rey_arXiv_2013,Zhang_arXiv_2014}. The spin-spin couplings arise from the direct overlap of single-particle wavefunctions, and because these systems must be dilute to avoid rapid two-body losses \cite{Bishof2011b,Ludlow2011} typical spin-spin interaction energies are fairly small. However, the long-lived nature of the optically excited  state, together with excellent coherence properties of state-of-the-art clock lasers \cite{Boyd_Science_2006,bloom:optical_2014,hinkley:atomic_2013}, both make coherent spin-dynamics accessible in these systems \cite{swallows:suppression_2011,lemke:p-wave_2011,martin:quantum_2013,Rey_arXiv_2013}. If $p$-wave interactions are ignored, the spin-spin Hamiltonian is SU$(2)$ symmetric ($J_z=J_{\perp}$), and no dynamics results from the collective initial states considered in this
manuscript (inhomogeneous states do, however, have non-trivial dynamics \cite{Campbell_Science_2009,rey:many-body_2009,gibble:decoherence_2009,yu:clock_2010}). The $p$-wave interactions induce an anisotropy $\tilde{J}=J_z-J_{\perp}$ which, while small, can still induce nontrivial dynamics (i.e., dynamics not described by short-time perturbation theory) for the collective initial states considered here, during timescales longer than  hundreds of ms.
 
\subsection{Quantum magnetism in tilted lattices}

The method, proposed in Ref.~\cite{Sachdev_PRB_2002} and realized in Refs.~\cite{Simon_Nature_2011,Meinert2013}, leaves behind the typical approach of hyperfine-encoded spin states with superexchange mediated interactions. The key idea is that the low energy manifold of a unit-filled Mott insulator, in the presence of a linear potential gradient, can be mapped onto the low-energy sector of a nearest-neighbor antiferromagnetic Ising model ($J_{\perp}=0$, $J_z>0$) in a transverse field. The smallest energy scale of the spin model is ultimately constrained by the tunneling energy in the lattice, which is about an order of magnitude larger than the superexchange scale that usually governs quantum magnetism of neutral atoms in optical lattices. The non-equilibrium dynamics described in this manuscript is only valid in the low tipping angle limit, because the mapping from a Bose Hubbard model to the antiferromagnetic Ising model is only valid within the low energy sectors of \emph{both} models. This does 
not necessarily mean, however, that the universal aspects of the dynamics we considered cannot be explored for weak quenches, where the dynamics is indeed governed by Eq.~\eqref{eq:Ising-model} plus an additional transverse field. Numerical simulations of the time evolution \cite{Rubbo2011},  that compared  the full  Bose Hubbard model  and the spin model  dynamics, after a quench  from a pure spin polarized state did show agreement for  various oscillations, thus confirming the validity of  this observation.

\section{Conclusions and outlook\label{sec:conclusions}}

In this paper we have studied the dynamics of spin systems  governed  by a general XXZ Hamiltonian  Eq.~\eqref{eq:XXZ-model}.  
In particular, we studied the time-evolution of an initially spin-polarized state, a protocol that can be viewed and experimentally implemented either as a quantum quench or as Ramsey spectroscopy. 
 We found that the range and magnitude of correlations and entanglement out of equilibrium could become comparable to or larger than those exhibited by strongly-correlated equilibrium ground states. In particular, in one dimension, we found that entanglement grows to satisfy a ``volume law" over time. The steady state nevertheless is not maximally entangled. We also demonstrated that universal singular dynamics could manifest out of equilibrium.

To derive these results, we employed a variety of exact analytic and numerical methods. Necessarily, the exact methods were restricted to special cases: short times, the Ising limit ($J^\perp_{ij}=0$), and one-dimensional systems. 
Alternative methods --- even approximate ones--- to study the more general cases are highly desirable. Among those one can mention:  mean-field theories, the truncated Wigner approximation~\cite{polkovnikov:phase_2010}, cluster expansions~\cite{witzel:quantum_2005,witzel:quantum_2006,maze:electron_2008,hazzard:many-body_2014}, and linear response theory ~\cite{hazzard:spectroscopy_2012}. We expect that the exact results developed here will provide a foundation and testbed for those approximate methods.

Many additional intriguing aspects of the physics are opened up by slight modifications of the dynamical procedure, several of which may be experimentally implementable. For example, one could change the sudden quenches to quenches with a finite, variable rate and thereby explore Kibble-Zurek type physics~\cite{kibble_topology_1976,zurek_cosmological_1985}. One could also work with inhomogeneous initial spin states  opening the possibility to study  transport and other  intriguing phenomena such as many-body localization~\cite{basko:many-body_2006,pal:many-body_2010,
pekker:hilbert-glass_2013,kwasigroch:bose-einstein_2013,yao:many-body_2013}, especially in systems with inhomogeneous and long-range couplings $J_{ij}^{\{\perp,z\}}$.

\section*{Acknowledgements}

We gratefully acknowledge Erez Berg, John Bollinger, Joe Britton, Vadim Cheianov, Jacob Covey, Eugene Demler, Bryce Gadway, Alexey Gorshkov, Murray Holland, Debbie Jin, Stefan Kehrein, Mikhail Lukin, Dominic Meiser, Steven Moses, Brian Neyenhuis, Brian Sawyer, Johannes Schachenmayer, Jon Simon, Michael Wall, Bo Yan,  Jun Ye, and Bihui Zhu for discussions around the work in this manuscript.

This work was supported by NIST, the NSF (PIF-1211914 and PFC-1125844), AFOSR and ARO individual investigator awards, and the ARO with funding from the DARPA-OLE program. K.~H. and M.~F.~F. thank the NRC for support, the Aspen Center for Physics, which is supported by the NSF, for its hospitality during the initial conception of this work, and the University of G{\"o}ttingen  for its hospitality during the final preparation of this manuscript. E.~D.~T. acknowledges the financial support of the Harvard-MIT CUA. T.~P. acknowledges support by the ERC under contract number 267100.
We thank  the Kavli Institute for Theoretical Physics (KITP) at UCSB, supported by NSF grant NSF PHY11-25915, for its hospitality while part of this work was carried out. This work utilized the Janus supercomputer, which is supported by the NSF (award number CNS-0821794) and the University of Colorado Boulder, and is a joint effort with the University of Colorado Denver and the National Center for Atmospheric Research. M.~K. acknowledges support by the National Research Foundation of South Africa under the Incentive Funding and the Competitive Program for Rated Researchers.

\bibliography{spin-noneq-correlation-submit-candidate}

\end{document}